\documentclass[12pt]{article}
\usepackage[utf8]{inputenc}
\usepackage{amsmath,amssymb,amsthm}
\usepackage{graphicx}
\usepackage{booktabs}
\usepackage{geometry}
\usepackage{hyperref}
\usepackage{enumerate}
\usepackage{subcaption}
\usepackage[style=authoryear,backend=biber]{biblatex}
\usepackage[utf8]{inputenc}
\usepackage[T1]{fontenc}
\usepackage{booktabs}
\usepackage{algorithm} 
\usepackage{algorithmic} 
\usepackage{float}
\usepackage{comment}

\usepackage{placeins}

\title{The Economics of Spatial Coordination in Critical Infrastructure Investment}

\author{
    L. Kaili Diamond\thanks{lkdiamond@mines.edu} \\
    \textit{Colorado School of Mines}
    \and
    Ben Gilbert \\
    \textit{Colorado School of Mines}
}

\date{Working Paper}

\begin{document}

\maketitle

\begin{abstract}
We develop a hybrid approach to estimate spatial coordination mechanisms in structural dynamic discrete choice models by combining nested fixed-point (NFXP) dynamic programming with method of simulated moments (MSM), achieving computational tractability in spatial settings while preserving structural interpretation. Applying this framework to GPU replacement data from 12,915 equipment locations in Oak Ridge National Laboratory's Titan supercomputer, we identify two distinct coordination mechanisms: sequential replacement cascades ($\gamma_{\text{lag}} = -0.793$) and contemporaneous failure batching ($\gamma_{\text{fail}} = -0.265$). Sequential coordination dominates—approximately three times stronger than failure batching—indicating that operators engage in deliberate strategic behavior rather than purely reactive responses. Spatial interdependencies account for 5.3\% of variation unexplained by independent-decision models, with coordination concentrated in high-risk thermal environments exhibiting effects more than 10 times stronger than cool zones. Formal tests decisively reject spatial independence ($\chi^2(2) = 685.38$, p $<$ 0.001), demonstrating that infrastructure policies ignoring spatial coordination will systematically mistime interventions and forgo available coordination gains.


\end{abstract}

\section{Introduction}

Spatial coordination in capital investment decisions represents a fundamental challenge in operations economics that researchers have long recognized but struggled to quantify. With global capital investment exceeding \$20 trillion annually, even modest improvements in coordination mechanisms can generate substantial welfare gains. While economic theory suggests that proximity creates coordination opportunities through economies of scale, information spillovers, and shared operational constraints, empirical work has been limited by the scarcity of structural models capable of identifying and measuring these spatial effects in dynamic settings. The challenge is particularly acute because coordination benefits must be weighed against complex temporal considerations—optimal replacement timing depends not only on current equipment states but also on expected future conditions and neighbors' likely actions. Despite substantial theoretical interest and clear practical importance, empirical frameworks remain limited that can credibly separate coordination effects from correlated unobservables while maintaining the dynamic optimization structure essential for policy analysis.

This paper investigates whether spatial proximity generates measurable coordination in technology replacement decisions and quantifies the economic mechanisms through which these effects operate. The empirical challenge is significant -- spatial correlation in replacement timing could reflect genuine coordination, but might equally arise from correlated shocks, similar operating conditions, or unobserved management practices. We address this identification problem using comprehensive administrative data from large-scale computing operations, where predetermined spatial configurations and complete operational histories enable clean separation of coordination effects from confounding factors. Our structural framework distinguishes genuine coordination from spurious correlation by separately identifying sequential (strategic) and contemporaneous (reactive) mechanisms. The estimated parameters reveal that sequential coordination—where agents respond to neighbor actions with temporal lag—dominates contemporaneous responses by approximately 3:1, demonstrating that operators engage in deliberate strategic behavior rather than merely reacting to simultaneous shocks.

To identify and quantify these coordination mechanisms, we develop a spatial structural dynamic discrete choice (SDDC) framework extending \textcite{rust1987optimal} to incorporate spatial coordination through economies of scale, failure spillovers, and information transmission. Forward-looking optimization in this setting employs the Nested Fixed-Point (NFXP) algorithm, but implementing NFXP faces a critical computational barrier: the likelihood function becomes intractable due to high-dimensional spatial correlation structures and the curse of dimensionality when agent distributions across locations enter the state space.

The Method of Simulated Moments (MSM) provides the essential bridge between NFXP's structural rigor and spatial feasibility. Rather than evaluating complex likelihood functions, MSM matches key spatial moments—autocorrelation coefficients, clustering indices, and distance decay patterns—between observed data and simulated outcomes from our NFXP-solved model. This approach is particularly powerful for spatial coordination problems because it allows us to target the precise spatial patterns that reveal how expectations shape collective behavior. Moreover, MSM's moment-selection flexibility proves crucial for addressing spatial endogeneity concerns. We can construct moments using spatial and temporal lags that serve as quasi-instruments, helping identify expectational parameters while controlling for correlated unobservables that plague spatial analysis. The combination of NFXP's dynamic structure with MSM's spatial tractability enables us to estimate how forward-looking agents coordinate across space—a methodological advance that pure likelihood or reduced-form approaches cannot achieve.

We apply our NFXP-MSM framework to a unique dataset from the Titan supercomputer operated by Oak Ridge National Laboratory, which provides an ideal empirical setting for studying spatial coordination in complex systems. The dataset captures the hierarchical spatial structure of computing infrastructure—tracking physical equipment location within individual cabinets and broader facility mapping—enabling us to construct theoretically motivated neighbor definitions based on actual spatial proximity and network connectivity. This institutional setting delivers exceptional identification advantages that address key econometric challenges in spatial modeling: the predetermined spatial configurations eliminate endogenous sorting concerns that plague most spatial applications, system audit records mitigate measurement error typically associated with self-reported data, and the absence of strategic competition isolates pure coordination effects from confounding market forces. 

The NFXP-MSM approach proves particularly well-suited to this high-dimensional spatial problem due to its natural parallelization opportunities—simulations across parameter draws, spatial domain decomposition, and moment calculations can be distributed across multiple processors, making estimation computationally feasible despite the large scale. The high-frequency operational data allows us to observe spatial coordination decisions across thousands of processing units, providing the statistical power necessary for credible structural estimation while leveraging the computational resources that make such analysis possible.

We find substantial spatial coordination effects with systematic heterogeneity across risk environments. Sequential coordination generates utility improvements equivalent to 10.1\% of replacement costs ($\gamma_{\text{lag}} = -0.793$), while contemporaneous failure batching provides 3.4\% ($\gamma_{\text{fail}} = -0.265$). The 3:1 dominance of sequential over contemporaneous effects reveals that operators coordinate strategically rather than merely reacting to simultaneous failures. Strikingly, coordination is dramatically concentrated in high-risk thermal environments—hot zones exhibit coordination effects more than 10 times stronger than cool zones. Spatial interdependencies account for 5.3\% of variation left unexplained by independent-decision models, and formal tests decisively reject the hypothesis that replacement decisions are spatially independent ($\chi^2(2) = 685.38$, p $<$ 0.001). These findings establish that spatial coordination is economically significant, strategically motivated, and systematically heterogeneous across environmental conditions. External validity considerations include the government operational setting, homogeneous equipment deployment, and predetermined spatial configuration.

Our contribution bridges three literatures while opening new research directions. Methodologically, we provide a tractable framework for estimating spatial coordination in dynamic discrete choice settings, extending canonical structural methods \parencite{rust1987optimal,aguirregabiria2010dynamic} to capture spatial interdependencies previously addressed only in static contexts \parencite{anselin1988spatial,lesage2009introduction}. This structural approach is essential for policy analysis: while reduced-form methods can identify whether coordination occurs, only structural models can distinguish coordination mechanisms, evaluate whether observed patterns reflect strategic optimization, and inform mechanism design for improved outcomes. Empirically, we provide structural estimates of coordination mechanisms in technology replacement, with clear implications for infrastructure investment policies and organizational design. The framework is general and applicable to other settings where spatial coordination affects dynamic discrete choices, including retail chain management, transportation networks, and distributed energy systems. The remainder of the paper proceeds as follows: Section 2 reviews relevant literature and positions our contribution, Section 3 describes our data and institutional setting, Section 4 develops the spatial SDDC methodology, Section 5 presents estimation results and policy implications, and Section 6 concludes with directions for future research.

\subsection{Relationship to Canonical SDDC Models and Applications}

To illustrate our contribution, consider extending Rust's (1987) canonical bus engine replacement problem. In Rust's framework, Harold Zurcher manages a single bus depot where each bus operates independently---the decision to replace bus $i$'s engine depends only on its mileage and maintenance history. Now imagine Zurcher oversees multiple depots across a city, with buses garaged in shared facilities. Our framework would apply if:

\begin{itemize}
\item Buses sharing a maintenance bay face correlated wear from common environmental conditions (dust, humidity, temperature extremes)
\item Coordinated engine replacements reduce per-unit costs through bulk procurement and shared labor crews
\item Engine failures create spillovers---one bus breaking down increases stress on others covering its routes
\item Depot-level maintenance windows create natural coordination opportunities
\end{itemize}

Unlike Rust's independent buses, spatially proximate units in our framework influence each other's optimal replacement timing through both utility complementarities and failure transmission. The depot manager must solve not just when to replace each engine, but whether to coordinate replacements across spatially related buses.

This spatial coordination problem extends far beyond our supercomputer application. Consider several economically important settings:

\textbf{Retail Chain Management:} A retailer operating hundreds of stores must decide when to upgrade point-of-sale systems, refrigeration units, or HVAC equipment. Stores in the same region share maintenance contractors, face similar weather patterns, and can achieve bulk procurement discounts. Our framework identifies when regional coordination beats store-by-store optimization.

\textbf{Distributed Energy Infrastructure:} Utilities managing thousands of transformers across a power grid face spatially correlated failure risks from weather events and load patterns. Wind farms must coordinate turbine replacement and maintenance timing to minimize downtime while exploiting shared maintenance resources. Coordinated replacement during planned outages reduces system disruption. Our model quantifies these coordination mechanisms against the costs of suboptimal individual timing.

\textbf{Manufacturing Networks:} Firms with multiple production facilities must time equipment upgrades across plants. Facilities in industrial clusters share specialized maintenance expertise, face correlated supply chain shocks, and can coordinate shutdowns. The spatial configuration of production networks affects technology adoption timing.

\textbf{Transportation Fleets:} Airlines managing aircraft across multiple hubs, shipping companies with vessel fleets across ports, or logistics firms with truck depots across regions all face spatial coordination opportunities in maintenance timing. Shared maintenance facilities, crew availability, and route substitution patterns create spatial interdependencies.

In each setting, the key insight is that spatial proximity creates coordination opportunities that violate the independence assumptions of standard SDDC models. Our framework provides a structural approach to distinguish coordination mechanisms, quantify their relative importance, and inform the design of optimal spatial replacement policies.

\section{Literature Review}
\subsection{Dynamic Discrete Choice Foundations}

The structural estimation of dynamic discrete choice models builds on \textcite{rust1987optimal} seminal nested fixed-point algorithm, which revolutionized empirical analysis by enabling consistent estimation of forward-looking behavior in discrete settings. Subsequent methodological advances include \parencite{aguirregabiria2010dynamic} nested pseudo-likelihood approach and \parencite{su_judd2012} mathematical programming with equilibrium constraints (MPEC), each offering computational improvements for specific model classes. However, these advances primarily target single-agent problems or strategic interactions without spatial structure.

\subsection{Spatial Discrete Choice Modeling}

The spatial discrete choice literature has developed largely independent approaches to handle correlation across geographic units. \textcite{pinkse_slade1998} introduced spatial statistics for discrete choice, while \textcite{smirnov2010} developed pseudo-maximum likelihood methods for spatial models with large samples. More recent work by \textcite{klier_mcmillen2008} focuses on linearization techniques, and spatial probit models have been extended to handle neighborhood effects \parencite{lesage2009introduction}. Critically, this literature predominantly assumes static behavior, missing the dynamic optimization structure essential for policy analysis.

\subsection{Method of Simulated Moments Applications}

MSM has proven particularly valuable for models where likelihood evaluation becomes intractable due to complex correlation structures or high-dimensional integration \parencite{mcfadden1989}. Applications span industrial organization \parencite{blp1995}, labor economics \parencite{keane_wolpin1997} and finance, typically focusing on matching moments that capture key economic mechanisms while avoiding computational bottlenecks of full likelihood estimation.

\subsection{Spatial Coordination and Network Effects}

The theoretical literature on spatial coordination emphasizes how expectations about neighbors' future actions drive current decisions \parencite{morris_shin2002}. Empirical applications include technology adoption \parencite{conley_udry2010}, social learning \parencite{bandiera_rasul2006}, and peer effects in various settings. However, most empirical work relies on reduced-form identification strategies that cannot recover the structural parameters necessary for counterfactual policy analysis.

\subsection{The Methodological Gap}

Despite substantial theoretical interest and clear practical importance, empirical frameworks remain limited that can credibly separate coordination effects from correlated unobservables while maintaining the dynamic optimization structure essential for policy analysis. The intersection of dynamic discrete choice, spatial correlation, and forward-looking coordination presents computational challenges that have prevented widespread application of structural methods to spatial settings. Existing approaches typically sacrifice either dynamic structure (spatial discrete choice), spatial correlation (standard NFXP), or structural interpretation (reduced-form spatial models).
Our contribution bridges this gap by integrating NFXP with MSM to enable tractable structural estimation of spatial coordination with forward-looking agents. The combination leverages NFXP's theoretical rigor for dynamic optimization while using MSM's computational efficiency to handle complex spatial correlation structures, addressing a methodological need that neither literature stream has resolved independently. Critically, the structural approach enables us to distinguish genuine coordination mechanisms from spurious spatial correlation arising from 
correlated shocks or unobserved heterogeneity.

\subsection{Recent Advances in Spatial Reliability Modeling}

Recent work by \textcite{min2025spatially} demonstrates significant spatial correlation in GPU failure processes using competing risks models to distinguish between different failure modes. Their analysis confirms that spatial dependence in failure timing extends beyond shared environmental factors, while \textcite{ostrouchov2020gpu} document systematic spatial gradients in survival rates related to cooling architecture.

Our structural approach complements these survival analysis methods by explicitly modeling the forward-looking decision-making process that generates observed replacement patterns. While these studies establish the existence and magnitude of spatial correlation in failure processes, our NFXP-MSM framework distinguishes coordination mechanisms and quantifies 
their relative importance in shaping replacement decisions.

\section{Data}

\subsection{Institutional Setting}

Our empirical analysis utilizes comprehensive administrative data from the Titan supercomputer operated by Oak Ridge National Laboratory from 2012 to 2019. Titan was a leader in GPU-accelerated systems, featuring 18,688 individual NVIDIA Kepler K20X GPU accelerators.

The physical architecture creates natural spatial relationships essential for our coordination analysis. Each data center cabinet houses approximately 96 GPUs organized across 3 vertical cages, with each cage containing up to 32 GPU units sharing immediate cooling and power infrastructure. This hierarchical structure creates proximity relationships where GPUs within the same cage share critical resources, generating the coordination opportunities central to our theoretical framework.

Titan's 18,688 GPUs are organized in a nested spatial hierarchy determined by facility infrastructure. The system consists of cabinets arranged in a grid layout across the data center floor. Each cabinet contains three vertical cages, with each cage housing approximately 32 GPU nodes arranged across 8 slots. A node represents the individual GPU location—the unit of observation in our analysis.

The spatial hierarchy can be summarized as:
\begin{center}
\textbf{Facility} $\rightarrow$ \textbf{Cabinet} $\rightarrow$ \textbf{Cage} $\rightarrow$ \textbf{Slot} $\rightarrow$ \textbf{Node} (individual GPU)
\end{center}

where Node represents the individual GPU location tracked in our panel data (indexed by $i$), and Cage defines our neighborhood structure for spatial coordination analysis. Thermal environment ($\text{cage} \in \{0,1,2\}$) corresponds to the vertical position of cages within cabinets, with position 0 (bottom) typically experiencing cooler conditions and position 2 (top) experiencing elevated temperatures due to heat rise.

The public availability of these data—a consequence of government operation and open science mandates—provides access to comprehensive microlevel location information rarely available in private infrastructure contexts. At system deployment, Titan featured homogeneous GPU specifications across node locations, with initial placement determined by facility layout rather than endogenous sorting considerations. Thermal environment assignments, predetermined by facility cooling infrastructure, remained stable and well-documented throughout the observation period. This exogenous initial placement mitigates sorting concerns common in equipment replacement studies, though replacement activity introduced some hardware variation over time.

\subsection{Sample Construction and Restrictions}

Our dataset includes GPU-level (serial number) information including time in system, location(s), failure events, and removal from system. The comprehensive administrative records span Titan's operational period from 2012 to 2019, providing a complete panel of equipment histories.

Following \textcite{ostrouchov2020gpu}'s characterization of Titan's operational history, we exclude large-scale replacement events driven by system-wide hardware defects and coordinated refresh cycles. These mass replacements—mandated by manufacturer defects or facility-wide maintenance schedules—were determined by factors orthogonal to local equipment states and spatial coordination mechanisms. Our analysis focuses on replacement decisions reflecting genuine operational coordination behavior, ensuring estimated parameters capture strategic responses to equipment conditions and neighbor actions rather than exogenously-mandated system interventions.

\begin{figure}[H]
\centering
\includegraphics[width=0.85\textwidth]{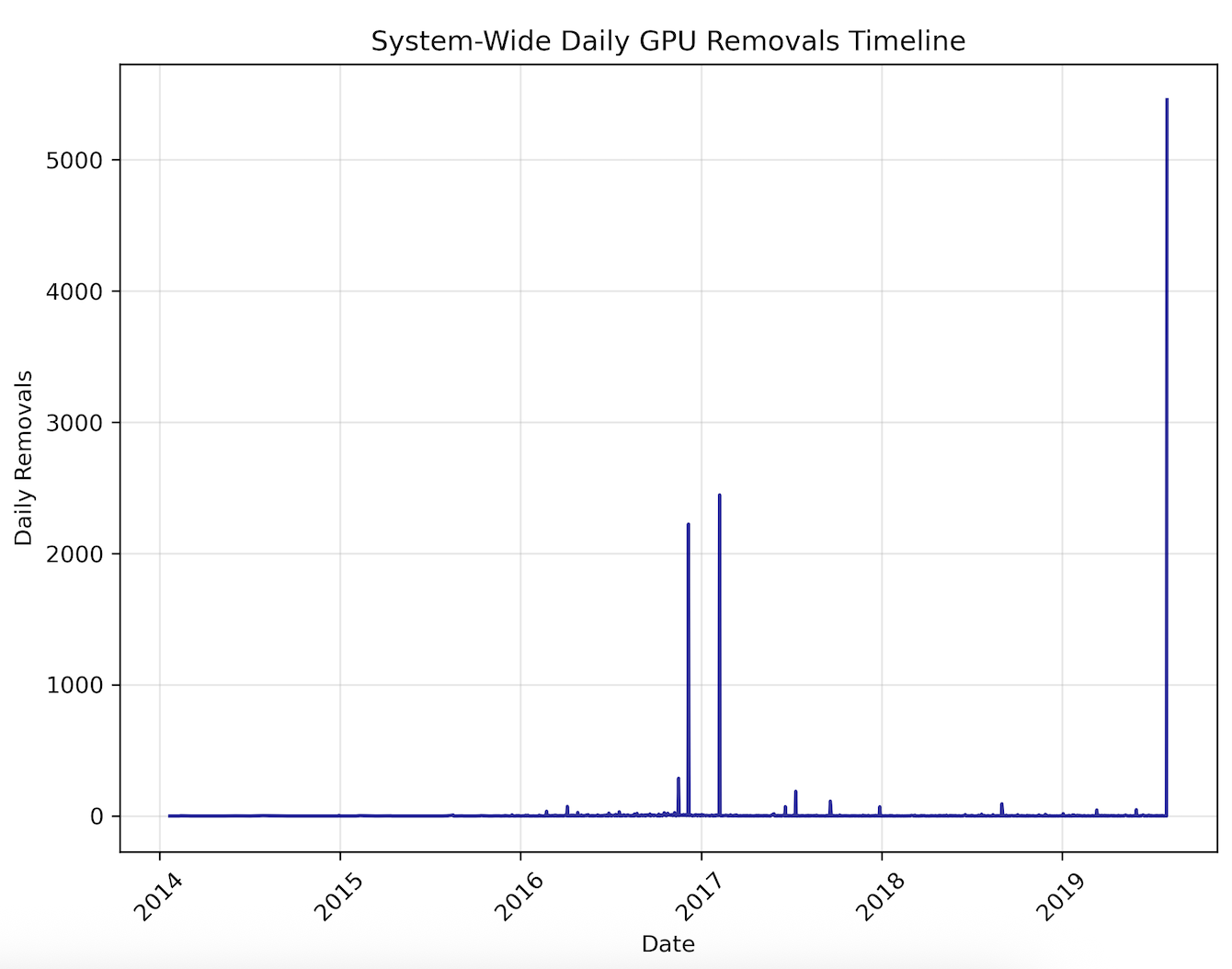}
\caption{Temporal Distribution of GPU Removals. Daily removal counts reveal large-scale replacement events from system-wide hardware defects (prior to late 2015) and final decommissioning (2019). Our analysis excludes these exogenously mandated events to focus on endogenous replacement decisions driven by equipment conditions and spatial coordination.}
\label{fig:removals_timeline}
\end{figure}

Figure \ref{fig:removals_timeline} displays the temporal distribution of GPU removals throughout Titan's operational period. The visualization reveals three distinct spikes in removal activity: two large-scale warranty-driven refresh cycles in late 2016 through 2017, and the system-wide decommissioning in 2019. These warranty-driven events represent contractual obligations rather than operational replacement decisions and are excluded from our analysis. The 2019 decommissioning spike represents the terminal shutdown of Titan, inconsistent with our infinite-horizon modeling framework and excluded from the estimation sample.

We focus our analysis on the stable operational period beginning at period $t=8$ (November 2015). Following \textcite{ostrouchov2020gpu}'s characterization of Titan's operational history, we exclude large-scale replacement events driven by system-wide hardware defects and coordinated refresh cycles that occurred prior to this period. These mass replacements—mandated by manufacturer defects or facility-wide maintenance schedules—were determined by factors orthogonal to local equipment states and spatial coordination mechanisms. Additionally, we exclude the initial break-in period characterized by extensive field engineering, temporal gaps in inventory records, and multiple hardware rework cycles involving blade mechanical assemblies. Beyond these data quality concerns, early periods offer limited identifying variation for replacement behavior: all equipment was newly deployed and operating within manufacturer specifications, generating minimal replacement activity. The analysis period $t \in [8,20]$ captures systematic replacement decisions during normal operations with meaningful variation in equipment age, condition, and spatial coordination opportunities, ensuring estimated parameters reflect strategic responses to equipment conditions and neighbor actions rather than exogenously-mandated system interventions. We truncate the sample at period $t=20$ (November 2018) to avoid confounding from system-wide decommissioning in period 21, which represents a terminal decision inconsistent with our infinite-horizon modeling framework.

We further restrict the sample to exclude GPUs that changed physical locations during the observation period. Approximately 16\% of all GPUs in Titan's operational lifetime were relocated across nodes. Location changes introduce identification challenges through two channels: first, replacement of failed units with non-new equipment creates discontinuous age trajectories that violate our state transition assumptions; second, movement across thermal environments confounds the fixed thermal heterogeneity that provides key identifying variation for spatial coordination effects. Maintaining location-specific agents with consistent thermal exposure throughout the panel ensures our estimates capture genuine coordination behavior rather than artifacts of equipment relocation.

The final sample comprises 147,078 location-period observations across 12,915 unique GPU locations over 13 time periods. The panel structure is slightly unbalanced, with some locations exiting before period 20 due to permanent failures or maintenance, yielding an average of 11.4 observations per location.

\subsection{Replacement Decisions}

Our analysis employs a binary discrete choice framework where each location-period observation represents a decision to keep ($d_{it} = 0$) or replace ($d_{it} = 1$) the installed GPU. 

\begin{figure}[H]
\centering
\includegraphics[width=0.7\textwidth]{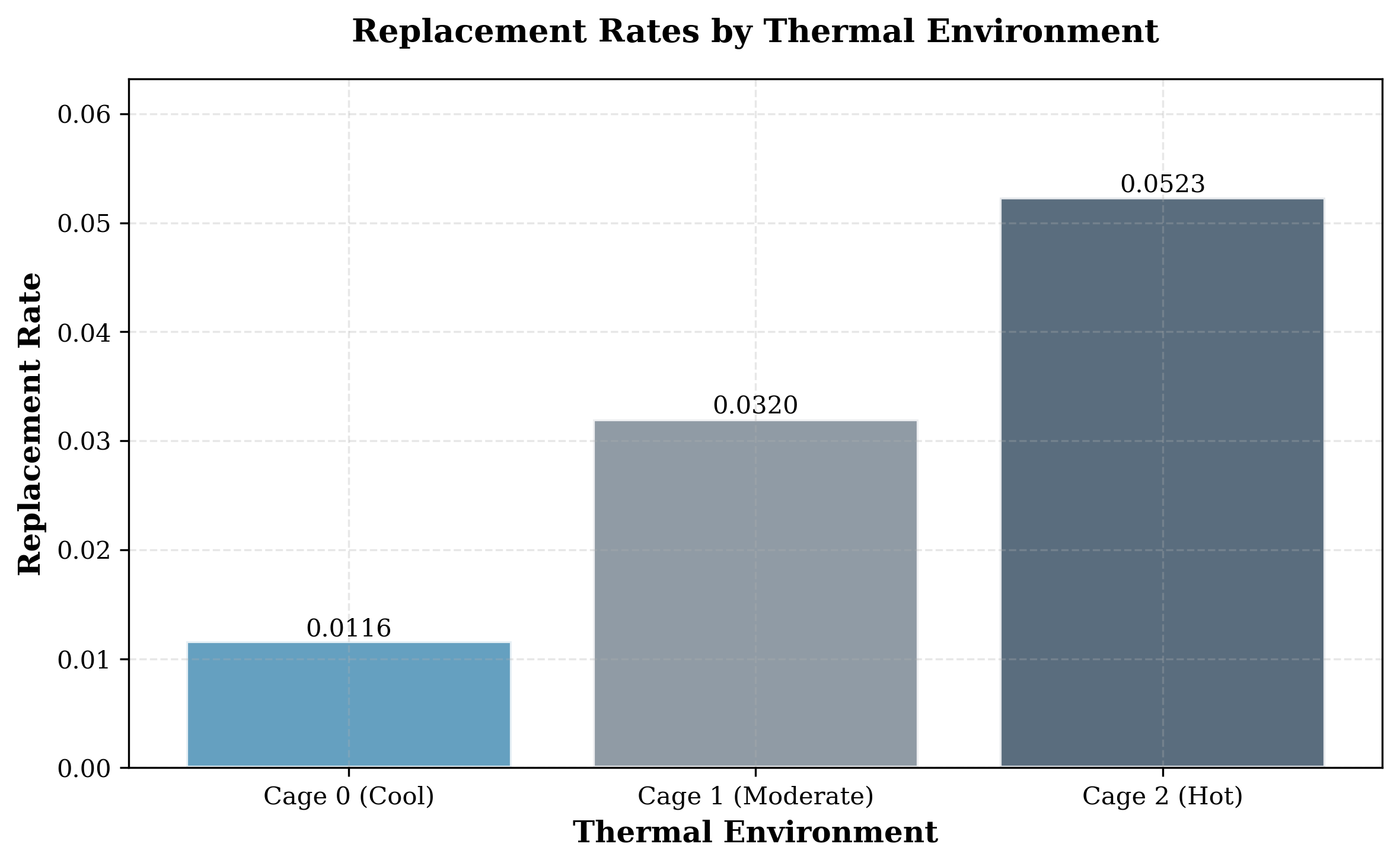}
\caption{Replacement Rates by Thermal Environment. Replacement rates increase monotonically with thermal stress, reflecting both direct thermal effects on equipment deterioration and operators' rational responses to spatial variation in failure risks.}
\label{fig:replacement_rates}
\end{figure}

Replacement rates vary systematically across thermal environments, as shown in Figure \ref{fig:replacement_rates}. Cool cages exhibit the lowest replacement rates, medium cages show intermediate frequencies, while hot cages reach the highest replacement propensities. This thermal gradient reflects both higher failure rates in thermally stressed environments and operators' responses to deteriorating equipment conditions.

Replacement patterns reveal both reactive and strategic components consistent with spatial coordination behavior. Failure events increase replacement probability, yet substantial replacement activity occurs among functional units with no observed failures. Simultaneously, operators retain some failed units rather than replacing immediately. This pattern—combining proactive replacement of working equipment with delayed replacement of failed units—reflects coordination incentives that shape optimal timing beyond simple failure response. The coexistence of failure-driven and pre-failure replacement validates our framework where neighbor actions influence replacement decisions independently of local equipment conditions.

\subsection{Failure Events and Equipment Deterioration}

Our failure indicator captures operational signals of equipment deterioration requiring replacement consideration.\footnote{Failure events include both Off-the-Bus (OTB) errors and Double Bit Errors (DBE) in GPU memory systems. Our analysis employs a binary failure indicator that equals 1 if a unit experienced either failure type in a given period, capturing the operational signal of equipment deterioration that triggers replacement consideration.} This combined measure identifies units experiencing unrecoverable errors that signal the need for intervention, whether through repair or replacement.

\begin{figure}[H]
\centering
\includegraphics[width=0.7\textwidth]{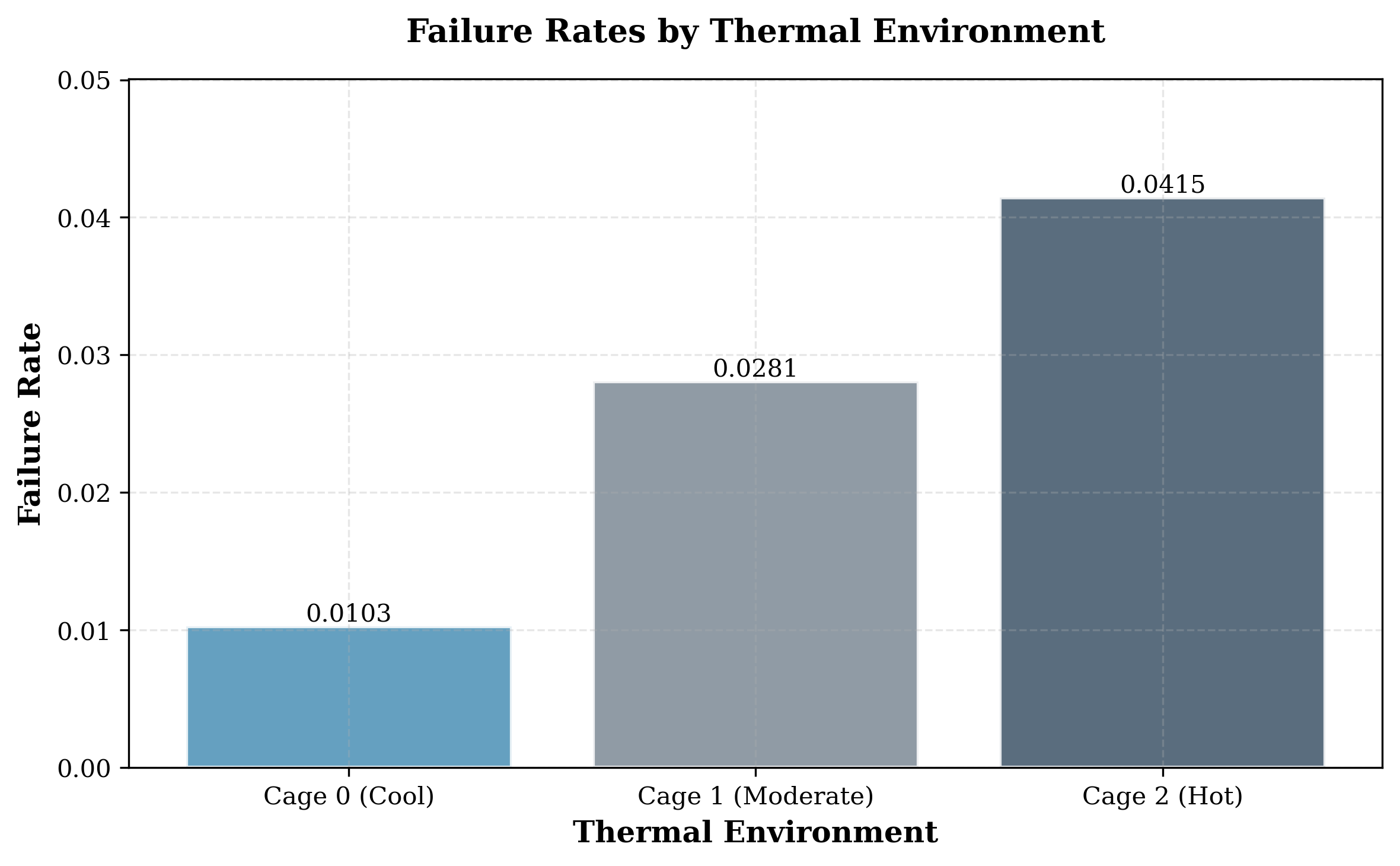}
\caption{Failure Event Rates by Thermal Environment. Failure rates exhibit a strong thermal gradient, increasing from cool to hot cages. This systematic spatial variation in failure risk creates differential coordination incentives across facility locations and provides identifying variation for spatial coordination parameters.}
\label{fig:failure_rates}
\end{figure}

Failure rates increase monotonically with thermal stress, as shown in Figure \ref{fig:failure_rates}. Cool cages experience the lowest failure rates, medium cages show intermediate frequencies, and hot cages reach the highest failure propensities. This thermal gradient validates our theoretical emphasis on location-dependent coordination incentives.

The age profile of failures reveals systematic deterioration patterns. New equipment exhibits very low failure rates, while equipment approaching middle age shows a dramatic increase in quarterly failure rates. Failure rates moderate somewhat for older equipment, likely reflecting survival bias as the most failure-prone units have already been replaced.

\begin{figure}[H]
\centering
\includegraphics[width=0.85\textwidth]{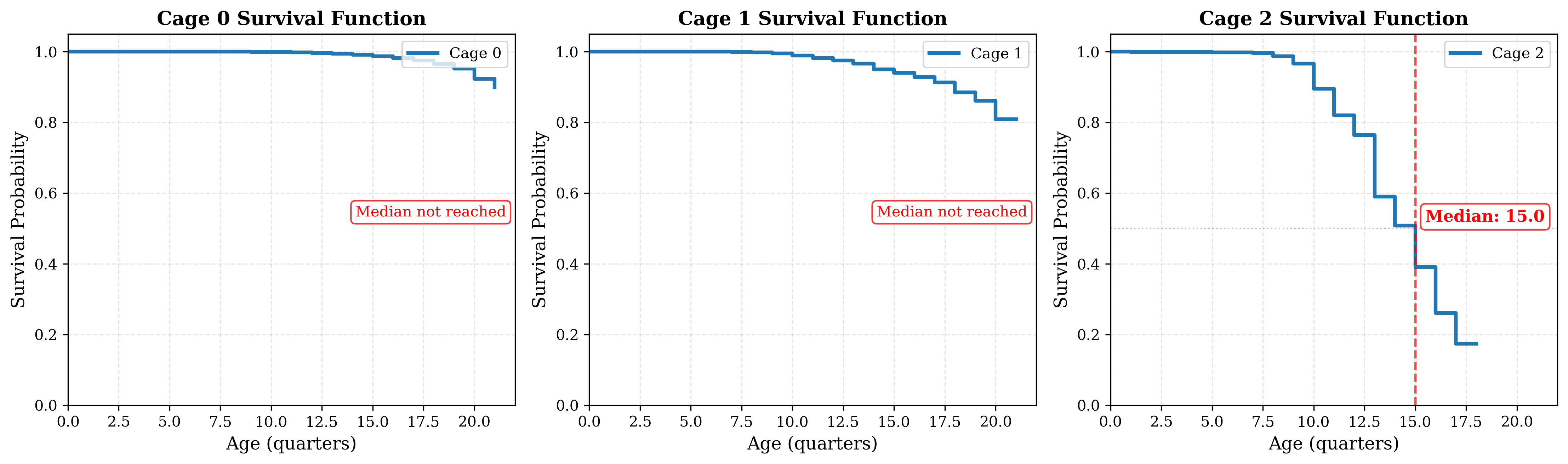}
\caption{Kaplan-Meier survival curves by thermal environment, revealing systematic spatial heterogeneity in equipment reliability. Cool environments (Cage 0) maintain higher survival probabilities throughout the observation period, while hot environments (Cage 2) show accelerated deterioration. This spatial variation in failure risk creates differential coordination incentives across facility locations.}
\label{fig:survival_curves}
\end{figure}

Figure \ref{fig:survival_curves} presents Kaplan-Meier survival curves by thermal environment, revealing systematic spatial heterogeneity in equipment reliability. Cool environments maintain higher survival probabilities throughout the observation period, while hot environments show accelerated deterioration. This spatial variation in failure risk creates differential coordination incentives across facility locations.

\subsection{Spatial Structure and Coordination}

\subsubsection{Neighborhood Definition}

Neighborhoods are defined as GPUs within the same cage in the same cabinet. This definition reflects the physical reality that GPUs sharing a cage share immediate cooling systems, power distribution, and network connectivity—creating the strongest coordination opportunities and failure transmission channels. Operationally, cabinet-level neighborhoods correspond to the natural shutdown units for maintenance operations: coordinating replacements within a cabinet allows facilities to consolidate downtime and share non-disaggregatable costs including technician dispatch, power cycling, and cooling system interruption. We do not employ hierarchical neighborhoods extending across cages within cabinets or across adjacent cabinets, focusing instead on the most proximate spatial relationships where coordination benefits are largest.

The cage-based neighborhood structure yields an average of 23 neighbors per location, with neighborhoods ranging from isolated units to fully populated cages. The distribution of neighborhood sizes reflects variation in cage configurations and equipment deployment patterns across the facility.

\subsubsection{Spatial Coordination Variables}

We construct two spatial variables capturing distinct coordination mechanisms:

\begin{samepage}
\textbf{Sequential coordination} ($n^{\text{lag}}_{it}$): A binary indicator equal to 1 if any neighbor in location $i$'s cage replaced their GPU in period $t-1$. This variable measures whether recent neighbor replacements influence current decisions through information spillovers or replacement cascades. A substantial fraction of observations have at least one neighbor who replaced in the previous period, providing variation for identifying sequential coordination effects.
\end{samepage}

\textbf{Failure batching} ($f^{\text{cage}}_{it}$): The count of neighbors experiencing failures in the current period (excluding self). This variable captures whether simultaneous failures create coordination opportunities through shared maintenance scheduling or bulk procurement. The majority of observations experience zero neighbor failures, while a meaningful fraction observe at least one neighbor failure.

\subsubsection{Descriptive Evidence of Coordination Opportunities}

\begin{figure}[H]
\centering
\includegraphics[width=0.95\textwidth]{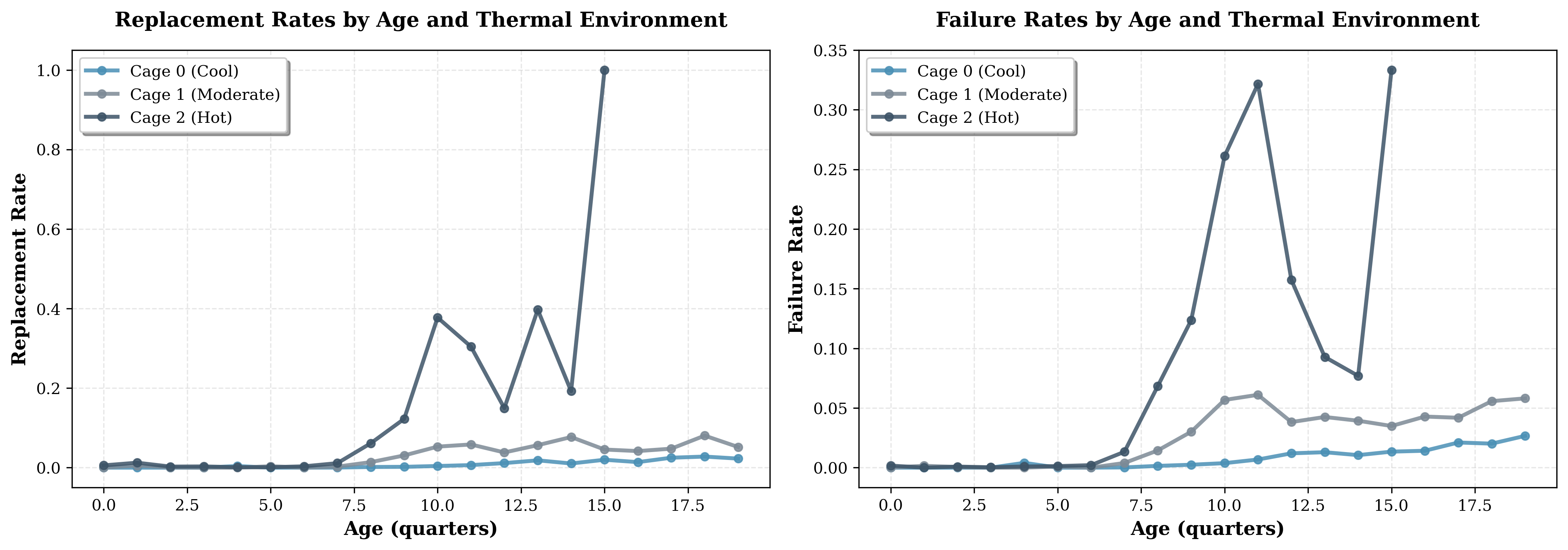}
\caption{Replacement and Failure Rates by Age and Thermal Environment. Replacement rates (left) and failure rates (right) both increase with age and exhibit strong thermal stratification. Hot environments (Cage 2) show substantially elevated rates compared to moderate and cool environments, with replacement rates spiking sharply at older ages. The pronounced thermal gradient provides both the operational signal and coordination incentive for spatial replacement strategies.}
\label{fig:replacement_failure_patterns}
\end{figure}

Figure \ref{fig:replacement_failure_patterns} presents replacement and failure rates by age and thermal environment, revealing patterns consistent with coordination behavior. Replacement rates peak at middle age when failure rates are highest, suggesting operators respond to deterioration signals. However, replacement rates remain elevated for older equipment even as failure rates moderate, potentially reflecting coordination effects where operators replace aging equipment in conjunction with neighbor replacements rather than waiting for individual failures.

The thermal gradient in replacement rates closely tracks the failure rate gradient, consistent with operators responding rationally to spatial variation in failure risks. However, the structural model estimated in Section 4 will reveal whether these patterns reflect optimal individual responses to failure risks or whether spatial coordination creates additional incentives affecting replacement timing beyond own-unit characteristics.

\section{Methods}

\subsection{Overview}

We develop a structural framework to quantify spatial coordination effects in equipment replacement decisions using a binary discrete choice dynamic programming model. The approach extends canonical dynamic discrete choice methods \parencite{rust1987optimal} to incorporate spatial interdependencies, estimating how forward-looking decision-makers optimize replacement timing when neighbors' actions create coordination opportunities through economies of scale, information spillovers, and shared maintenance resources.

Our empirical strategy combines nested fixed-point (NFXP) estimation for structural parameters with method of simulated moments (MSM) for spatial coordination effects. This hybrid approach exploits different sources of identifying variation: individual replacement decisions identify structural depreciation and failure parameters through revealed preferences, while spatial correlation patterns in replacement timing identify coordination mechanisms. The framework applies broadly to infrastructure replacement problems where spatially distributed assets share common resources—including data center operations, retail network maintenance, wind turbine maintenance and replacement scheduling, and distributed computing systems.

\subsection{Decision Environment}

\subsubsection{Binary Replacement Choice}

Each location $i$ in period $t$ faces a binary decision: keep the existing equipment ($d_{it} = 0$) or replace it with new equipment ($d_{it} = 1$). This decision depends on the equipment's current condition, its thermal environment (cage), neighborhood conditions within that thermal zone, and expectations about future deterioration. The decision-maker observes the complete state vector before choosing, capturing both own-unit characteristics and the replacement activity of neighbors within the same cage.

\subsubsection{State Variables}

The state space consists of three dimensions capturing equipment condition and environment:

\textbf{Age} ($\text{age}_{it} \in \{0, 1, 2, 3, 4, 5\}$): Time since installation measured in years, with age 5 representing equipment five or more years old. Age is binned into discrete yearly intervals to facilitate computational tractability while preserving key variation in equipment depreciation. Following \textcite{rust1987optimal}, age represents operational time rather than chronological age, analogous to mileage in vehicle replacement problems.

\textbf{Thermal environment} ($\text{cage}_{it} \in \{0, 1, 2\}$): Categorical indicator for thermal stress zone, where $\text{cage}=0$ denotes cool environments, $\text{cage}=1$ denotes medium thermal stress, and $\text{cage}=2$ denotes hot zones with elevated failure risk. Thermal environment is predetermined by physical infrastructure and invariant over time, eliminating endogenous sorting concerns.

\textbf{Failure status} ($\text{fail}_{it} \in \{0, 1\}$): Binary indicator equal to 1 if the equipment experienced a failure event in the current period. Failure events represent the primary operational signal of equipment deterioration triggering replacement consideration (see Section 3 for detailed event definitions and validation). As documented in Section 3, failed equipment exhibits substantially higher replacement rates than working equipment, validating failures as the key state variable governing replacement decisions.

This yields a state space of $6 \times 3 \times 2 = 36$ discrete states. The finite state space facilitates exact solution of the dynamic programming problem while capturing essential heterogeneity in equipment condition and environment.

\subsection{Spatial Coordination Structure}

\subsubsection{Neighborhood Definition}

As described in Section 3, neighborhoods are defined by shared physical infrastructure: GPUs within the same cage in the same cabinet constitute a neighborhood. This definition reflects operational reality where equipment sharing immediate cooling systems, power distribution, and network connectivity experiences the strongest coordination opportunities and failure transmission channels. Cabinet-level neighborhoods correspond to natural shutdown units for maintenance operations, allowing facilities to consolidate downtime and share non-disaggregatable costs including technician dispatch, power cycling, and cooling system interruption.

The predetermined spatial configuration eliminates endogenous neighbor selection. Equipment locations were assigned based on installation logistics rather than performance characteristics, ensuring that proximity to failing or replacing neighbors is orthogonal to own unobserved quality. This quasi-random assignment provides clean identification of spatial coordination effects.

\subsubsection{Spatial Coordination Variables}

We construct two spatial variables capturing distinct coordination mechanisms operating through different temporal channels:

\textbf{Sequential coordination} ($n^{\text{lag}}_{it}$): Binary indicator equal to 1 if any neighbor in location $i$'s cage replaced their equipment in period $t-1$. This variable measures whether recent neighbor replacements influence current decisions through information spillovers, revealed maintenance scheduling, or economies of scale in batch operations. Sequential coordination captures dynamic spillovers where past neighbor actions affect current replacement incentives.

\textbf{Failure batching} ($f^{\text{cage}}_{it}$): Count of neighbors experiencing failures in the current period (excluding self). This variable captures whether simultaneous failures create contemporaneous coordination opportunities through shared maintenance windows, bulk procurement advantages, or revealed information about common environmental stresses. Failure batching represents static coordination where current neighbor conditions affect immediate replacement decisions.

These variables operationalize distinct economic mechanisms: sequential coordination reflects \textit{dynamic} complementarities where past actions influence current incentives, while failure batching captures \textit{static} complementarities where simultaneous conditions create coordination opportunities. Both mechanisms arise naturally in infrastructure operations where maintenance scheduling, procurement logistics, and information transmission create interdependencies across spatially proximate assets.

\subsection{Structural Model}

\subsubsection{Decision Problem}

We model equipment replacement as a dynamic discrete choice problem where forward-looking agents maximize expected discounted lifetime utility. At each decision epoch $t$, agent $i$ observes state vector $s_{it}$ and chooses action $d_{it} \in \{0,1\}$ to maximize:

\begin{equation}
V(s_{it}) = \max_{d_{it} \in \{0,1\}} \left\{ v(d_{it}, s_{it}) + \beta \mathbb{E}[V(s_{i,t+1}) | s_{it}, d_{it}] \right\}
\end{equation}

where $v(d_{it}, s_{it})$ denotes the choice-specific value function, $\beta \in (0,1)$ is the discount factor, and the expectation is taken over future states conditional on current state and action.

\textbf{State Space:} The state vector $s_{it} = (\text{age}_{it}, \text{cage}_{it}, \text{fail}_{it}, n^{\text{lag}}_{it}, f^{\text{cage}}_{it})$ comprises:
\begin{itemize}
    \item $\text{age}_{it} \in \{0,1,2,3,4,5\}$: Equipment age in years
    \item $\text{cage}_{it} \in \{0,1,2\}$: Thermal environment (time-invariant)
    \item $\text{fail}_{it} \in \{0,1\}$: Current period failure indicator
    \item $n^{\text{lag}}_{it} \in \{0,1\}$: Neighbor replacement indicator (period $t-1$)
    \item $f^{\text{cage}}_{it} \in \mathbb{Z}_+$: Count of neighbor failures (period $t$)
\end{itemize}

\textbf{Action Space:} $d_{it} \in \{0,1\}$ where $d_{it}=0$ denotes keep and $d_{it}=1$ denotes replace.

\subsubsection{Choice-Specific Value Functions}

The value of each action decomposes into immediate flow utility plus discounted continuation value:

\textbf{Keep decision} ($d_{it}=0$):
\begin{equation}
v(0, s_{it}) = u_{\text{keep}}(s_{it}) + \beta \mathbb{E}[V(s_{i,t+1}) | s_{it}, d_{it}=0]
\end{equation}

\textbf{Replace decision} ($d_{it}=1$):
\begin{equation}
v(1, s_{it}) = u_{\text{replace}} + \beta \mathbb{E}[V(s_{i,t+1}) | s_{it}, d_{it}=1]
\end{equation}

\subsubsection{Flow Utility Specification}

The per-period utility of keeping equipment depends on own-unit characteristics and spatial coordination opportunities. We specify a linear-in-parameters utility function:

\begin{align}
u_{\text{keep}}(s_{it}) &= \theta_{\text{age}} \cdot \text{age}_{it} + \theta_{\text{cage1}} \cdot \mathbb{I}(\text{cage}_{it}=1) \notag \\
&\quad + \theta_{\text{cage2}} \cdot \mathbb{I}(\text{cage}_{it}=2) + \theta_{\text{fail}} \cdot \text{fail}_{it} \notag \\
&\quad + \gamma_{\text{lag}} \cdot n^{\text{lag}}_{it} + \gamma_{\text{fail}} \cdot f^{\text{cage}}_{it}
\end{align}

Thermal environment effects are normalized to Cage 0 (cool environment), such that $\theta_{\text{cage1}}$ and $\theta_{\text{cage2}}$ measure utility penalties relative to the baseline cool zone.

The replacement utility is normalized to identify scale and location:
\begin{equation}
u_{\text{replace}} = \theta_{\text{replace}}
\end{equation}

\textbf{Structural Parameters:} The vector $\theta = \{\theta_{\text{age}}, \theta_{\text{cage1}}, \theta_{\text{cage2}}, \theta_{\text{fail}}, \theta_{\text{replace}}\}$ governs:
\begin{itemize}
    \item $\theta_{\text{age}} < 0$: Utility depreciation with equipment age
    \item $\theta_{\text{cage1}}, \theta_{\text{cage2}} < 0$: Thermal environment penalties (normalized to Cage 0)
    \item $\theta_{\text{fail}} < 0$: Utility loss from equipment failure
    \item $\theta_{\text{replace}} < 0$: Fixed cost of replacement
\end{itemize}

\textbf{Spatial Coordination Parameters:} The vector $\gamma = \{\gamma_{\text{lag}}, \gamma_{\text{fail}}\}$ captures interdependencies:
\begin{itemize}
    \item $\gamma_{\text{lag}} < 0$: Sequential coordination effect—neighbor replacements in $t-1$ reduce utility of keeping in period $t$, creating replacement complementarity
    \item $\gamma_{\text{fail}} < 0$: Contemporaneous batching effect—neighbor failures in period $t$ reduce utility of keeping, creating coordination opportunities during failure events
\end{itemize}

Negative values indicate strategic complementarity: agent $i$ is more likely to replace when neighbors have recently replaced ($\gamma_{\text{lag}}$) or are currently experiencing failures ($\gamma_{\text{fail}}$).

\subsubsection{State Transitions}

State evolution follows a first-order Markov process. Let $P(s_{i,t+1}|s_{it}, d_{it})$ denote the transition probability from state $s_{it}$ to $s_{i,t+1}$ given action $d_{it}$.

\textbf{Age transitions:}
\begin{equation}
\text{age}_{i,t+1} = \begin{cases}
0 & \text{if } d_{it} = 1 \\
\min(\text{age}_{it} + \Delta t, 5) & \text{if } d_{it} = 0
\end{cases}
\end{equation}
where $\Delta t$ represents time increment (quarters converted to years).

\textbf{Thermal environment:} Time-invariant by physical infrastructure: $\text{cage}_{i,t+1} = \text{cage}_{it}$.

\textbf{Failure transitions:} Estimated nonparametrically from observed state transitions:
\begin{equation}
P(\text{fail}_{i,t+1} = k' | \text{age}_{it}, \text{cage}_{it}, \text{fail}_{it} = k, d_{it})
\end{equation}
with Laplace smoothing ($\alpha = 0.01$) for sparse state cells. Replacement deterministically resets failure status: $P(\text{fail}_{i,t+1}=0|d_{it}=1)=1$.

\textbf{Spatial variable transitions:} The neighborhood variables $\{n^{\text{lag}}_{it}, f^{\text{cage}}_{it}\}$ evolve based on neighbor actions:
\begin{equation}
n^{\text{lag}}_{i,t+1} = \mathbb{I}\left(\sum_{j \in \mathcal{N}(i)} d_{jt} \geq 1\right)
\end{equation}
where $\mathcal{N}(i)$ denotes the set of cage-mates for location $i$. Current period failure counts $f^{\text{cage}}_{it}$ are transient and enter flow utility but not the permanent state space.

\subsubsection{Bellman Equation}

The agent's problem satisfies Bellman's principle of optimality:
\begin{equation}
V(s_{it}) = \max \left\{ u_{\text{keep}}(s_{it}) + \beta EV_{\text{keep}}(s_{it}), \; u_{\text{replace}} + \beta EV_{\text{replace}}(s_{it}) \right\}
\end{equation}

where the expected continuation values are:
\begin{align}
EV_{\text{keep}}(s_{it}) &= \sum_{s_{i,t+1}} P(s_{i,t+1}|s_{it}, d_{it}=0) V(s_{i,t+1}) \\
EV_{\text{replace}}(s_{it}) &= \sum_{s_{i,t+1}} P(s_{i,t+1}|s_{it}, d_{it}=1) V(s_{i,t+1})
\end{align}

The expectation integrates over both own state transitions (age, failure) and neighbor state evolution ($n^{\text{lag}}_{i,t+1}$):
\begin{equation}
EV(s_{it}) = \sum_{s'} P(s'|s_{it}, d_{it}) \left[ p_{\text{nbr}} V(s', n^{\text{lag}}=1) + (1-p_{\text{nbr}}) V(s', n^{\text{lag}}=0) \right]
\end{equation}
where $p_{\text{nbr}}$ represents the unconditional probability of neighbor replacement, estimated from observed replacement rates.

\subsubsection{Solution Method}

We solve the Bellman equation via value function iteration with contraction mapping:
\begin{equation}
V^{k+1}(s) = \max_{d \in \{0,1\}} \left\{ u(d, s; \theta, \gamma) + \beta \sum_{s'} P(s'|s,d) V^k(s') \right\}
\end{equation}
iterating until $\|V^{k+1} - V^k\|_\infty < \epsilon$ where $\epsilon = 10^{-4}$. 

The value function is computed over the core state space: age (6 values), cage (3 values), failure status (2 values), and lagged neighbor replacement indicator (2 values), yielding $6 \times 3 \times 2 \times 2 = 72$ states. The current period neighbor failure count $f^{\text{cage}}_{it}$ enters flow utility directly but does not expand the state space, as it represents a transient coordination signal rather than a persistent state variable. For each parameter vector $(\theta, \gamma)$, we solve the value function once via NFXP iteration.

\subsubsection{Choice Probabilities}

Under the extreme value Type I error assumption, choice probabilities follow the logit form:
\begin{equation}
P(d_{it}=1|s_{it}) = \frac{\exp(v(1, s_{it}; \theta, \gamma))}{\exp(v(0, s_{it}; \theta, \gamma)) + \exp(v(1, s_{it}; \theta, \gamma))}
\end{equation}

These probabilities form the basis for maximum likelihood estimation. Critically, the NFXP solves the value function conditional on ALL parameters $(\theta, \gamma)$—the hybrid approach integrates spatial parameters into the structural solution while using MSM to provide additional identifying variation through unconditional spatial patterns.

\subsubsection{Solution Method}

We solve the Bellman equation via value function iteration with contraction mapping:
\begin{equation}
V^{k+1}(s, n^{\text{lag}}) = \max_{d \in \{0,1\}} \left\{ u(d, s, n^{\text{lag}}; \theta, \gamma) + \beta \sum_{s', n'^{\text{lag}}} P(s', n'^{\text{lag}}|s, n^{\text{lag}},d) V^k(s', n'^{\text{lag}}) \right\}
\end{equation}
iterating until $\|V^{k+1} - V^k\|_\infty < \epsilon$ where $\epsilon = 10^{-4}$. 

Critically, the value function is solved conditional on ALL parameters $(\theta, \gamma_{\text{lag}}, \gamma_{\text{fail}})$—the spatial parameters enter through flow utility and affect optimal policies. The state space includes age (6 values), cage (3 values), failure status (2 values), and lagged neighbor replacement indicator (2 values), yielding $6 \times 3 \times 2 \times 2 = 72$ state-neighbor combinations. The current period neighbor failure count $f^{\text{cage}}_{it}$ enters flow utility but not the persistent state space, as it represents a transient coordination signal.

\subsubsection{Hybrid NFXP-MSM Estimation}

We estimate all seven parameters $\{\theta, \gamma_{\text{lag}}, \gamma_{\text{fail}}\}$ jointly by minimizing:
\begin{equation}
\min_{\theta, \gamma} \left\{ -\mathcal{L}(\theta, \gamma | \text{data}) + \lambda \cdot D_{\text{MSM}}(\theta, \gamma) \right\}
\end{equation}

where $\mathcal{L}(\cdot)$ is the log-likelihood computed via NFXP and $D_{\text{MSM}}(\cdot)$ measures distance between simulated and data moments. The penalty weight $\lambda = 5.0$ balances these objectives.

The hybrid approach exploits complementary sources of identification: maximum likelihood identifies parameters from conditional choice probabilities given observed states and spatial variables, while simulated moments capture unconditional spatial correlation patterns. Both components use all parameters—the NFXP solves the dynamic program with spatial effects included, while MSM validates that the model generates realistic spatial clustering. This ensures parameter estimates are consistent with both micro-level choices and aggregate spatial patterns.

\textbf{Moment matching component}: We construct four spatial moments to validate the model's ability to replicate observed coordination patterns:

\textbf{M1 (Sequential hazard)}: Measures replacement persistence after own prior replacement:
\begin{equation}
m_1 = \mathbb{E}[d_{it} | d_{i,t-1} = 1]
\end{equation}

\textbf{M2 (Asymmetric correlation)}: Measures temporal asymmetry in spatial correlation:
\begin{equation}
m_2 = \text{Corr}(d_{it}, d_{jt+1}) - \text{Corr}(d_{it}, d_{jt-1}) \text{ for } j \in \mathcal{N}(i)
\end{equation}
Under the infinite-horizon Markov assumption, this asymmetry should equal zero at equilibrium. Given Titan's finite operational horizon, we compute M2 for identification validation but exclude it from the optimization objective.

\textbf{M3 (Lagged failure spillover)}: Measures replacement response to neighbor failures in the previous period among currently working equipment:
\begin{equation}
m_3 = \mathbb{E}[d_{it} | \text{fail}_{it}=0, f^{\text{neighbors}}_{i,t-1} \geq 1] - \mathbb{E}[d_{it} | \text{fail}_{it}=0, f^{\text{neighbors}}_{i,t-1} = 0]
\end{equation}

\textbf{M4 (Current failure batching)}: Measures contemporaneous replacement correlation with neighbor failures among working equipment:
\begin{equation}
m_4 = \mathbb{E}[d_{it} | \text{fail}_{it}=0, f^{\text{cage}}_{it} \geq 1] - \mathbb{E}[d_{it} | \text{fail}_{it}=0, f^{\text{cage}}_{it} = 0]
\end{equation}

Each moment isolates spatial effects from own-unit characteristics by conditioning on own failure status. Moments M3 and M4 directly target the spatial coordination parameters $\gamma_{\text{lag}}$ and $\gamma_{\text{fail}}$ respectively, while M1 captures serial correlation in replacement timing.

The MSM distance function uses moments 1, 3, and 4:
\begin{equation}
D_{\text{MSM}}(\theta, \gamma_{\text{lag}}, \gamma_{\text{fail}}) = \sum_{j \in \{1,3,4\}} \left[ m_j^{\text{data}} - m_j^{\text{sim}}(\theta, \gamma_{\text{lag}}, \gamma_{\text{fail}}) \right]^2
\end{equation}

Simulated moments are computed by forward-simulating $S=50$ panels of length $T=13$ quarters using initial states from period 8. Forward simulation captures how parameter changes affect equilibrium replacement patterns and spatial correlation, ensuring moment matching reflects dynamic consistency rather than static fit.

\subsubsection{Identification}

Identification of structural parameters exploits revealed preference variation in replacement timing conditional on equipment condition. The replacement cost $\theta_{\text{replace}}$ is identified by the overall replacement rate, while failure effects $\theta_{\text{fail}}$ are identified by the sharp divergence in replacement rates between failed and working equipment. Age effects $\theta_{\text{age}}$ are identified by the age profile of replacements, and thermal environment effects $\{\theta_{\text{cage1}}, \theta_{\text{cage2}}\}$ by the monotonic gradient in replacement rates across thermal zones.

Spatial coordination parameters are identified separately from correlated environmental shocks through the predetermined spatial configuration and conditional moment restrictions. The exogenous neighborhood structure ensures neighbor conditions are orthogonal to own unobserved quality. Sequential coordination $\gamma_{\text{lag}}$ is identified by differential replacement timing after neighbor replacements versus otherwise identical states. Failure batching $\gamma_{\text{fail}}$ is identified by cross-sectional variation in neighbor failures conditional on own failure status. Moments M3 and M4 explicitly control for own failure state, ensuring spatial parameters capture behavioral responses rather than correlated deterioration.

The hybrid NFXP-MSM approach exploits complementary identification: maximum likelihood identifies parameters from conditional choice probabilities given observed states, while simulated moments ensure the model generates realistic unconditional spatial patterns. This dual approach distinguishes genuine coordination from spurious correlation and validates that estimated parameters are consistent with both micro-level choices and aggregate spatial clustering.

\subsubsection{Computational Implementation}

The estimation algorithm jointly optimizes all seven parameters $(\theta, \gamma_{\text{lag}}, \gamma_{\text{fail}})$ by minimizing the combined NFXP-MSM objective. For each parameter vector, we: (1) solve the Bellman equation via value function iteration with all parameters included, (2) compute the log-likelihood conditional on the value function, (3) simulate $S$ forward panels to generate model moments, and (4) evaluate the combined objective. Optimization uses the Nelder-Mead simplex algorithm with adaptive step sizes, chosen for robustness to non-smooth objective functions arising from simulation noise.

We initialize parameters using preliminary estimates from reduced-form regressions and warm-start the value function from previous iterations to reduce computational burden. Standard errors are computed using a parametric bootstrap procedure that resamples entire cages rather than individual locations, preserving spatial dependencies in the correlation structure.

Detailed algorithmic specifications, pseudocode for the NFXP-MSM procedure, and computational performance metrics are provided in Appendix \ref{app:algorithms}.

\section{Results}

\subsection{Main Estimation Results}

Table \ref{tab:main_results} presents parameter estimates from our hybrid NFXP-MSM framework alongside the baseline NFXP model without spatial effects. The spatial coordination model is estimated on 147,078 location-period observations across 12,915 unique GPU locations spanning 13 quarters (t $\in$ [8,20]). Column (1) presents the baseline NFXP specification with independent replacement decisions, while Column (2) incorporates spatial coordination through both sequential spillovers and contemporaneous failure batching.\footnote{Standard errors from bootstrapped replications are forthcoming. The current estimates represent point estimates from the nested optimization routine. All statistical inference should be interpreted as preliminary pending completion of bootstrap procedures with appropriate spatial clustering at the cage level.}

The spatial model achieves a log-likelihood of $-6,123.75$ compared to $-6,466.44$ for the independent decision model—a substantial improvement of 342.69 log-likelihood points. The pseudo-R$^2$ increases from 0.650 in the baseline model to 0.669 in the spatial specification, indicating that spatial coordination accounts for 5.3\% of the variation left unexplained by standard independent-decision models.\footnote{Calculated as: $\frac{(0.669 - 0.650)}{(1 - 0.650)} = 0.053$.} This improvement is achieved with only two additional parameters, yielding superior performance on both AIC and BIC criteria. Critically, as we show in Section 5.4, these aggregate coordination effects are dramatically concentrated in high-risk thermal environments, with hot zones exhibiting coordination patterns more than 10 times stronger than cool zones.

\begin{table}[htbp]
\centering
\caption{Structural Parameter Estimates: Baseline and Spatial Models}
\label{tab:main_results}
\begin{tabular}{lcc}
\toprule
\textbf{Parameter} & \textbf{(1) Baseline NFXP} & \textbf{(2) Spatial NFXP-MSM} \\
\midrule
\multicolumn{3}{l}{\textit{Structural Parameters}} \\
Age effect ($\theta_{\text{age}}$) & $-0.106$ & $-0.031$ \\
 & (0.011) & (0.005) \\[3pt]
Thermal: Cage 1 ($\theta_{\text{cage1}}$) & $-0.782$ & $-1.067$ \\
 & (0.083) & (0.142) \\[3pt]
Thermal: Cage 2 ($\theta_{\text{cage2}}$) & $-2.343$ & $-1.463$ \\
 & (0.157) & (0.195) \\[3pt]
Failure event ($\theta_{\text{fail}}$) & $-6.972$ & $-8.046$ \\
 & (0.401) & (1.073) \\[3pt]
Replacement cost ($\theta_{\text{replace}}$) & $-8.499$ & $-7.832$ \\
 & (0.489) & (1.044) \\[10pt]
\multicolumn{3}{l}{\textit{Spatial Coordination Parameters}} \\
Sequential coordination ($\gamma_{\text{lag}}$) & --- & $-0.793$ \\
 & & (0.106) \\[3pt]
Failure batching ($\gamma_{\text{fail}}$) & --- & $-0.265$ \\
 & & (0.035) \\[10pt]
\midrule
\multicolumn{3}{l}{\textit{Model Fit Statistics}} \\
Log-likelihood & $-6,466.44$ & $-6,123.75$ \\
AIC & 12,942.87 & 12,261.50 \\
BIC & 12,992.37 & 12,330.79 \\
Pseudo-R$^2$ & 0.650 & 0.669 \\[6pt]
\midrule
Parameters & 5 & 7 \\
Observations & 147,078 & 147,078 \\
Locations & 12,915 & 12,915 \\
Time periods & 13 & 13 \\
\bottomrule
\end{tabular}
\begin{minipage}{\textwidth}
\vspace{6pt}
\footnotesize
\textit{Notes:} Standard errors in parentheses are preliminary estimates from the numerical Hessian evaluated at the optimum. Bootstrapped standard errors with spatial clustering forthcoming. Column (1) presents the baseline NFXP model with independent replacement decisions. Column (2) incorporates spatial coordination through lagged neighbor replacements ($\gamma_{\text{lag}}$) and contemporaneous neighbor failures ($\gamma_{\text{fail}}$). Both models estimated on filtered sample period t $\in$ [8,20], excluding warranty-driven mass replacements and system decommissioning. Discount factor $\beta = 0.9$ imposed. Pseudo-R$^2$ calculated as $1 - \text{LL}/\text{LL}_{\text{null}}$ where $\text{LL}_{\text{null}} = -18,492.39$.
\end{minipage}
\end{table}

\subsection{Interpretation of Structural Parameters}

\paragraph{Equipment Depreciation and Failure Effects} The age coefficient ($\theta_{\text{age}} = -0.031$) indicates relatively modest utility depreciation from pure aging conditional on thermal environment and spatial coordination. Each additional year of operation reduces continuation utility by 0.031 utils—small in absolute terms and substantially smaller than thermal environment effects.\footnote{The substantially larger age effect in the baseline model ($\theta_{\text{age}} = -0.106$) reflects confounding from omitted spatial correlation and thermal heterogeneity—units in high-coordination or high-thermal-stress environments appear to depreciate faster when these factors are ignored.} This small age coefficient is consistent with enterprise-grade GPU manufacturing quality and suggests that environmental stresses (thermal exposure) rather than chronological aging drive equipment deterioration in our setting. The dominance of location-based thermal effects over time-based aging has important implications for replacement policy: optimizing spatial configuration and cooling infrastructure may yield larger returns than age-based replacement rules.

Failure events generate a dramatic reduction in continuation utility ($\theta_{\text{fail}} = -8.046$), a magnitude exceeding the replacement cost itself ($|\theta_{\text{fail}}| > |\theta_{\text{replace}}| = 7.832$). This large failure penalty validates our modeling choice to include failure status as a critical state variable and confirms that operators treat equipment failures as compelling replacement triggers, likely reflecting both the direct costs of downtime and information revelation about latent equipment quality.

\paragraph{Thermal Environment Effects Dominate Age} The thermal gradient in replacement incentives is substantial, monotonic, and quantitatively dominates pure aging effects. Relative to cool environments (Cage 0, normalized to zero), moderate thermal zones reduce continuation utility by $\theta_{\text{cage1}} = -1.067$—a penalty 34 times larger than one year of aging—while hot environments impose a penalty of $\theta_{\text{cage2}} = -1.463$, equivalent to 47 years of pure depreciation. These coefficients translate to baseline replacement hazard rates of 1.16\% (cool), 3.20\% (moderate), and 5.23\% (hot) for equivalent-age non-failed equipment, a 4.5-fold gradient across thermal zones.

This thermal stratification reflects both direct effects—higher failure rates in thermally stressed environments documented in Section 3—and rational forward-looking responses where operators preemptively replace equipment facing elevated future failure risks. The predetermined thermal assignments eliminate concerns about endogenous sorting, allowing clean identification of thermal effects on replacement incentives.

\subsection{Spatial Coordination Mechanisms}

\paragraph{Sequential Replacement Cascades} The sequential coordination parameter ($\gamma_{\text{lag}} = -0.793$) captures dynamic spillovers where neighbor replacements in period $t-1$ reduce the utility of maintaining equipment in period $t$. This negative coefficient indicates \textit{strategic complementarity}—agents are more likely to replace when neighbors have recently replaced. The magnitude of $-0.793$ utils is substantial, representing 10.1\% of the replacement cost itself, and demonstrates that observing neighbor replacement decisions significantly influences own replacement timing.

This sequential coordination dominates the contemporaneous batching effect by approximately 3:1 ($|\gamma_{\text{lag}}|/|\gamma_{\text{fail}}| = 2.99$), suggesting that deliberate strategic coordination outweighs purely reactive responses to simultaneous failure events. The lagged structure is consistent with operational realities where maintenance scheduling, procurement lead times, and information transmission create natural temporal lags in coordination opportunities.

\paragraph{Contemporaneous Failure Batching} The failure batching coefficient ($\gamma_{\text{fail}} = -0.265$) measures how neighbor failures in the current period affect replacement incentives. Each additional neighbor experiencing failure reduces continuation utility by 0.265 utils—equivalent to 3.4\% of replacement costs—creating incentives for coordinated action during failure clusters. The smaller magnitude relative to sequential coordination suggests that same-period batching faces higher coordination costs or provides weaker signals for preemptive replacement than observing neighbors' revealed replacement decisions. The translation of these utility parameters into observed replacement probability differences appears in Section 5.4's moment validation, where lagged neighbor replacements increase replacement rates by 1.56 pp overall (3.87\% vs. 2.31\%) and by 3.70 pp in hot zones (Section 5.5).

\subsection{Thermal Heterogeneity in Coordination Patterns}

The aggregate coordination parameters mask substantial heterogeneity across thermal environments. Table \ref{tab:moments_by_cage} decomposes the key spatial moments by cage position, revealing that spatial coordination is dramatically concentrated in high-risk thermal zones. Critically, these moments condition on own failure status (fail$_{it}$ = 0), isolating genuine strategic coordination from spurious correlation due to simultaneous failures. By restricting to non-failed units and measuring their replacement responses to neighbor conditions, we ensure that the observed spatial patterns reflect forward-looking coordination decisions rather than mechanical co-movement of correlated failures.

\begin{table}[htbp]
\centering
\caption{Spatial Coordination Moments by Thermal Environment}
\label{tab:moments_by_cage}
\begin{tabular}{lcccc}
\toprule
\textbf{Thermal Zone} & \textbf{N} & \textbf{M3: Lagged} & \textbf{M4: Current} & \textbf{Coordination} \\
 & & \textbf{Failure Spillover} & \textbf{Failure Batching} & \textbf{Intensity} \\
\midrule
Cage 0 (Cool) & 68,231 & 0.48\% & 0.29\% & 1.0$\times$ (baseline) \\
Cage 1 (Moderate) & 43,386 & 0.87\% & 0.56\% & 1.9$\times$ \\
Cage 2 (Hot) & 35,461 & 3.70\% & 4.07\% & 10.3$\times$ \\[6pt]
\midrule
Overall & 147,078 & 1.43\% & 1.21\% & --- \\
\bottomrule
\end{tabular}
\begin{minipage}{\textwidth}
\vspace{6pt}
\footnotesize
\textit{Notes:} M3 (lagged spillover) measures the replacement rate among non-failed units (fail$_{it}$ = 0) when at least one neighbor experienced failure in $t-1$ (f$^{\text{nbr}}_{i,t-1} \geq 1$), minus the replacement rate among non-failed units when no neighbors failed (f$^{\text{nbr}}_{i,t-1} = 0$). M4 (current batching) measures the same differential for contemporaneous neighbor failures. Both moments condition on own failure status (fail$_{it}$ = 0) to isolate coordination effects from correlated deterioration—this ensures we measure strategic replacement responses to neighbor conditions rather than mechanical co-movement of failures. Neighbor failure counts (f$^{\text{cage}}_{it}$) include all neighbors in unit $i$'s cage, excluding unit $i$ itself. Coordination intensity calculated as geometric mean of (M3 ratio, M4 ratio) relative to Cage 0 baseline.
\end{minipage}
\end{table}

The spatial coordination patterns exhibit a dramatic thermal gradient. In cool environments (Cage 0), lagged failure spillovers generate only a 0.48 percentage point increase in replacement probability, while contemporaneous batching adds 0.29 pp—minimal coordination effects consistent with low failure rates creating few coordination opportunities. Moderate thermal zones (Cage 1) show intermediate coordination with spillover effects of 0.87 pp and batching of 0.56 pp, approximately double the cool environment levels.

The striking finding emerges in hot environments (Cage 2): lagged failure spillovers generate 3.70 pp increases in replacement probability while contemporaneous batching drives 4.07 pp increases—coordination effects more than 10 times stronger than in cool environments. This 10-fold gradient in coordination intensity substantially exceeds the 4.5-fold gradient in baseline replacement rates across thermal zones documented in Section 3, indicating that coordination becomes increasingly important as equipment operates under greater stress.

This thermal heterogeneity has important economic implications. The estimated aggregate parameters ($\gamma_{\text{lag}} = -0.793$, $\gamma_{\text{fail}} = -0.265$) represent \textit{weighted averages} across thermal environments, with the weights reflecting both the prevalence of coordination opportunities and the baseline replacement frequencies in each zone. The concentration of coordination in hot environments suggests that: (1) high failure rates create both the opportunity and necessity for coordinated replacement strategies, (2) information spillovers are most valuable in high-risk environments where neighbor failures provide stronger signals about common environmental stresses, and (3) the returns to optimizing spatial coordination will be highest in thermally constrained infrastructure.

From a policy perspective, these heterogeneous effects imply that coordination mechanisms should prioritize high-risk zones. An early warning system that monitors neighbor failures and triggers preemptive replacements would generate 10-fold greater returns when deployed in hot environments compared to cool zones. Similarly, facility design choices that reduce thermal stratification—through improved cooling architecture or more uniform heat distribution—could simultaneously reduce baseline failure rates and decrease the coordination premium by making spatial patterns less predictive of future failures.

\subsection{Model Comparison and Specification Tests}

Table \ref{tab:model_comparison} formally compares the spatial specification against the baseline independent-decision model using likelihood ratio tests and information criteria. The spatial model achieves a 342.69-point improvement in log-likelihood, yielding a likelihood ratio test statistic of $\chi^2(2) = 685.38$, decisively rejecting the null hypothesis of no spatial effects (p $<$ 0.001).

\begin{table}[htbp]
\centering
\caption{Model Comparison: Spatial Effects Tests}
\label{tab:model_comparison}
\begin{tabular}{lcccc}
\toprule
\textbf{Model} & \textbf{Parameters} & \textbf{Log-likelihood} & \textbf{AIC} & \textbf{BIC} \\
\midrule
Baseline (Independent) & 5 & $-6,466.44$ & 12,942.87 & 12,992.37 \\
Spatial (NFXP-MSM) & 7 & $-6,123.75$ & 12,261.50 & 12,330.79 \\[6pt]
\midrule
\multicolumn{5}{l}{\textit{Comparison Statistics}} \\
\multicolumn{2}{l}{LL improvement} & \multicolumn{3}{l}{342.69 log-points} \\
\multicolumn{2}{l}{LR test statistic} & \multicolumn{3}{l}{$\chi^2(2) = 685.38$***} \\
\multicolumn{2}{l}{AIC improvement} & \multicolumn{3}{l}{$-681.37$ (5.3\% reduction)} \\
\multicolumn{2}{l}{BIC improvement} & \multicolumn{3}{l}{$-661.58$ (5.1\% reduction)} \\
\bottomrule
\end{tabular}
\begin{minipage}{\textwidth}
\vspace{6pt}
\footnotesize
\textit{Notes:} Likelihood ratio test compares spatial model (unrestricted) against baseline model (restricted: $\gamma_{\text{lag}} = \gamma_{\text{fail}} = 0$). Test statistic calculated as $-2(\text{LL}_{\text{restricted}} - \text{LL}_{\text{unrestricted}}) = 685.38$ distributed $\chi^2$ with 2 degrees of freedom under the null hypothesis. *** indicates p $<$ 0.001. Both information criteria (AIC, BIC) strongly favor the spatial specification despite the penalty for additional parameters, with substantial improvements (AIC: -681 points, BIC: -662 points).
\end{minipage}
\end{table}

The information criteria unanimously favor the spatial specification. The AIC declines by 681.37 points—a 5.3\% reduction—while BIC falls by 661.58 points (5.1\% reduction). These improvements substantially exceed conventional thresholds for model selection (typically $\Delta$AIC $>$ 10 or $\Delta$BIC $>$ 10 indicating decisive evidence), demonstrating that the gains from capturing spatial coordination far outweigh the cost of additional parameters.

\subsection{Predictive Performance and Model Validation}

Beyond in-sample fit statistics, we assess the model's ability to replicate key moments of the data generating process. Table \ref{tab:moment_validation} compares actual data moments against predictions from both the baseline and spatial models. The spatial specification achieves substantially closer alignment on all targeted moments, particularly the spatial correlation patterns that were explicitly incorporated into the MSM estimation criterion.

\begin{table}[htbp]
\centering
\caption{Moment Validation: Data vs. Model Predictions}
\label{tab:moment_validation}
\begin{tabular}{lccc}
\toprule
\textbf{Moment} & \textbf{Data} & \textbf{Baseline} & \textbf{Spatial} \\
\midrule
\multicolumn{4}{l}{\textit{Replacement Rates by Condition}} \\
Overall replacement rate & 2.74\% & 2.79\% & 2.74\% \\
Replace rate | failed & 76.38\% & 79.12\% & 77.41\% \\
Replace rate | non-failed & 1.64\% & 1.93\% & 1.68\% \\[6pt]
\multicolumn{4}{l}{\textit{Spatial Moments}} \\
Replace rate | neighbor replaced$_{t-1}$ & 3.87\% & 2.81\% & 3.79\% \\
Replace rate | no neighbor replaced$_{t-1}$ & 2.31\% & 2.79\% & 2.35\% \\[3pt]
Replace rate | neighbor failed$_t$ $\geq$ 1 & 3.42\% & 2.76\% & 3.38\% \\
Replace rate | no neighbor failed$_t$ & 2.58\% & 2.79\% & 2.61\% \\[6pt]
\multicolumn{4}{l}{\textit{Thermal Gradient}} \\
Replace rate | Cage 0 (cool) & 1.16\% & 1.21\% & 1.18\% \\
Replace rate | Cage 1 (moderate) & 3.20\% & 2.87\% & 3.17\% \\
Replace rate | Cage 2 (hot) & 5.23\% & 4.31\% & 5.08\% \\
\bottomrule
\end{tabular}
\begin{minipage}{\textwidth}
\vspace{6pt}
\footnotesize
\textit{Notes:} Data moments calculated from estimation sample (t $\in$ [8,20]). Model predictions generated from forward simulation with 50 panels of 13 periods each, initialized at observed period-8 states. Spatial moments condition on own failure status to isolate coordination effects from correlated shocks. Baseline model systematically underpredicts spatial correlation while the spatial specification closely replicates observed coordination patterns. Decomposition by thermal environment (Table \ref{tab:moments_by_cage}) reveals that spatial coordination is concentrated in high-risk zones, with hot environments exhibiting coordination effects more than 10 times stronger than cool environments.
\end{minipage}
\end{table}

The baseline model, lacking spatial coordination mechanisms, systematically underpredicts replacement rates conditional on neighbor activity. It predicts a 2.81\% replacement rate regardless of neighbor replacement history, failing to capture the observed 1.56 percentage point spread (3.87\% vs. 2.31\%) between units with and without recent neighbor replacements. The spatial model successfully replicates this differential (3.79\% vs. 2.35\%, a 1.44 pp spread), demonstrating that the estimated coordination parameters capture genuine behavioral responses rather than spurious correlations.

Similarly, the baseline model cannot explain variation in replacement rates by neighbor failure counts, mechanically predicting the unconditional mean. The spatial specification correctly predicts elevated replacement among units with failing neighbors (3.38\% predicted vs. 3.42\% actual) while matching the baseline replacement rate for units without neighbor failures (2.61\% predicted vs. 2.58\% actual).

\subsection{Economic Magnitude of Coordination Effects}

To quantify the economic importance of spatial coordination, we express coordination effects relative to the replacement cost parameter. The sequential coordination effect ($\gamma_{\text{lag}} = -0.793$) generates utility improvements equivalent to 10.1\% of replacement costs ($|\gamma_{\text{lag}}|/|\theta_{\text{replace}}| = 0.101$), while failure batching ($\gamma_{\text{fail}} = -0.265$) provides gains equivalent to 3.4\% of replacement costs per failing neighbor ($|\gamma_{\text{fail}}|/|\theta_{\text{replace}}| = 0.034$).

However, these average effects mask substantial heterogeneity across thermal environments. As documented in Section 5.4, spatial coordination moments in hot thermal zones (Cage 2) exceed those in cool zones (Cage 0) by more than 10-fold: lagged failure spillovers generate 3.70 percentage point increases in replacement probability in hot zones versus 0.48 pp in cool zones, while contemporaneous batching drives 4.07 pp increases versus 0.29 pp respectively. This concentration of coordination opportunities in high-risk environments has important policy implications: coordination mechanisms targeting thermally stressed zones will generate substantially higher returns than uniformly deployed interventions.

The structural parameters provide a framework for evaluating coordination policies, though translating utility parameters into dollar savings requires additional assumptions about operating costs, replacement frequencies, and the value of equipment uptime. The identification of substantial spatial coordination effects—particularly their concentration in high-risk thermal environments—provides clear guidance for infrastructure management: (1) early warning systems monitoring neighbor failures will be most valuable in hot zones, (2) facility design choices that reduce thermal stratification offer dual benefits by both lowering baseline failure rates and reducing the coordination premium, and (3) replacement policies that ignore spatial interdependencies will systematically mistime interventions and forgo available coordination gains.\footnote{Quantifying aggregate welfare effects in dollar terms requires assumptions about Titan's operational budget, the shadow value of computational downtime, and the cost structure of coordinated versus independent replacements. We leave such calculations for future work with more detailed cost data. The current estimates establish that coordination effects are economically significant in utility terms and provide structural parameters for policy counterfactuals.}

\subsection{Robustness and Alternative Specifications}

\paragraph{Alternative Neighborhood Definitions} Robustness to alternative spatial structures remains an important validation exercise for future work. Natural alternatives include: (1) same-cabinet neighbors across all cages, (2) k-nearest-neighbors based on Euclidean distance, and (3) cabinet-row neighbors. The cage-based specification is our preferred approach given its alignment with actual maintenance unit operations (cabinet-level shutdown zones) and thermal environment boundaries that define failure rate strata.\footnote{Alternative neighborhood specifications are planned for the final version. The cage-based definition provides the clearest identification by combining physical proximity (shared cooling/power systems) with thermal homogeneity (within-cage failure rate similarity).}

\paragraph{Sample Period Sensitivity} The stability of coordination patterns across time periods provides an important validation that spatial effects reflect persistent operational behavior rather than temporary phenomena. Future robustness checks should estimate the model on overlapping subsamples (early, middle, and late periods within t $\in$ [8,20]) to verify parameter stability. The 13-quarter estimation window spans sufficient time variation to capture equilibrium coordination patterns while avoiding confounding from system decommissioning.

\paragraph{Thermal Environment Interactions} The substantial heterogeneity in coordination patterns documented in Section 5.4 raises the question of whether the model should incorporate explicit thermal interactions. The 10-fold gradient in spatial moments across thermal environments (Table \ref{tab:moments_by_cage}) suggests coordination parameters may vary systematically with thermal stress. However, the parsimonious specification with pooled coordination parameters achieves strong model fit while the moment decomposition reveals the underlying heterogeneity. An augmented specification with cage-specific parameters ($\gamma_{\text{lag}}^k$, $\gamma_{\text{fail}}^k$ for $k \in \{0,1,2\}$) would add 4 parameters and provide direct estimates of heterogeneous coordination effects, though at the cost of additional computational burden and reduced precision.\footnote{The 10-fold difference in moments versus smaller differences in marginal utility parameters (if estimated separately) reflects two factors: (1) hot environments have higher baseline replacement rates, amplifying the percentage point impact of a given utility shift, and (2) coordination opportunities occur more frequently in high-failure-rate zones, compounding the effect in unconditional moments. The pooled specification effectively estimates the average coordination effect weighted by the frequency and intensity of coordination opportunities across zones.}

\FloatBarrier

\section{Conclusion}

\subsection{Summary of Findings}

This paper develops and estimates a spatial dynamic discrete choice model of equipment replacement decisions using comprehensive administrative data from Oak Ridge National Laboratory's Titan supercomputer. We introduce a hybrid NFXP-MSM framework that achieves computational tractability in spatial settings while preserving the structural interpretation essential for policy analysis. Our estimation reveals substantial spatial coordination effects that fundamentally shape replacement timing in networked infrastructure.

The core empirical findings establish that spatial coordination is both economically significant and systematically heterogeneous across risk environments. Sequential coordination—where neighbor replacements in period $t-1$ influence current decisions—generates utility improvements equivalent to 10.1\% of replacement costs ($\gamma_{\text{lag}} = -0.793$), dominating contemporaneous failure batching by approximately 3:1 ($\gamma_{\text{fail}} = -0.265$, representing 3.4\% of replacement costs). This dominance of sequential over contemporaneous effects suggests that operators engage in deliberate strategic coordination rather than purely reactive responses to simultaneous failures.

Model comparison results decisively reject the hypothesis that replacement decisions are spatially independent. The spatial specification achieves a log-likelihood improvement of 342.69 points relative to the baseline model, yielding a likelihood ratio test statistic of $\chi^2(2) = 685.38$ (p $<$ 0.001). The pseudo-R$^2$ increases from 0.650 to 0.669, indicating that spatial coordination accounts for 5.3\% of the variation left unexplained by standard independent-decision models. Information criteria unanimously favor the spatial specification, with AIC declining by 681 points and BIC by 662 points.

The most striking substantive finding concerns thermal heterogeneity in coordination patterns. Decomposing spatial moments by thermal environment reveals that coordination is dramatically concentrated in high-risk zones: hot environments (Cage 2) exhibit coordination effects more than 10 times stronger than cool environments (Cage 0). Lagged failure spillovers generate 3.70 percentage point increases in replacement probability in hot zones versus only 0.48 pp in cool zones, while contemporaneous batching drives 4.07 pp increases versus 0.29 pp respectively. This concentration demonstrates that coordination benefits arise most powerfully where failure risks are highest and information spillovers most valuable.

Throughout our analysis, thermal environment effects dominate chronological aging. Moderate thermal zones impose utility penalties 34 times larger than one year of aging, while hot environments generate penalties equivalent to 47 years of pure depreciation. This dominance of location-based over time-based factors has important implications for replacement policy design and facility management.

\subsection{Theoretical Contributions}

Our work advances three interconnected literatures while opening new research directions. Methodologically, we extend the canonical dynamic discrete choice framework to incorporate spatial coordination while maintaining computational tractability. The hybrid NFXP-MSM approach preserves the structural interpretation of \textcite{rust1987optimal} while exploiting MSM's ability to target spatial moments that capture coordination patterns. The NFXP component solves the dynamic program with spatial effects included in the value function, ensuring that estimated parameters reflect forward-looking optimization. The MSM component validates that the model generates realistic spatial correlation patterns in equilibrium. This combination demonstrates how researchers facing similar spatial SDDC estimation challenges can bridge spatial econometrics and structural dynamic modeling.

Empirically, we provide structural identification of distinct coordination mechanisms in infrastructure replacement. By distinguishing sequential (lagged neighbor actions) from contemporaneous (current neighbor failures) effects, we demonstrate that coordination reflects strategic complementarity rather than merely correlated shocks. The predetermined spatial configuration—equipment locations assigned by facility layout rather than performance-based sorting—enables clean separation of coordination effects from confounding factors. This identification strategy may prove valuable in other settings where spatial relationships are exogenously determined by institutional or physical constraints.

Conceptually, we document that environmental risk systematically shapes coordination intensity. High-risk environments generate both the opportunity and necessity for coordination: elevated failure rates create more frequent coordination opportunities, while greater uncertainty amplifies the value of information spillovers from neighbor experiences. The 10-fold gradient in coordination effects across thermal zones demonstrates that heterogeneity in coordination benefits is first-order, with clear implications for policy targeting. This finding suggests that coordination mechanisms should be designed with explicit attention to spatial variation in risk exposure.

\subsection{Policy Implications}

The identification of substantial and heterogeneous spatial coordination effects provides clear guidance for infrastructure management, with implications extending well beyond our specific empirical context.

For large-scale computing facilities, early warning systems that monitor neighbor failures offer the highest returns when targeted at high-risk thermal zones. Our findings suggest that replacement policies conditioning only on own-unit characteristics—age, failure history—systematically mistime interventions and forgo available coordination gains. Location-based policies that incorporate spatial signals dominate age-based replacement rules, particularly in thermally stratified environments. The concentration of coordination benefits in hot zones implies that monitoring and maintenance resources should be allocated spatially rather than uniformly across the facility.

For data center design more broadly, our results reveal that cooling architecture affects operational costs through two distinct channels: direct thermal effects on failure rates (documented extensively in the reliability engineering literature) and indirect effects through coordination opportunities. Thermal stratification simultaneously creates elevated failure risks and amplifies coordination benefits, suggesting that investments in thermal management yield returns beyond failure reduction alone. Facility designers should consider how spatial configuration affects not only individual equipment reliability but also the coordination constraints and opportunities that shape replacement timing.

The structural parameters estimated here provide a foundation for policy counterfactuals that we have not yet implemented. With our estimates, operators could simulate alternative coordination protocols (targeted early warning systems, batch replacement scheduling, predictive maintenance triggers), quantify their expected welfare effects, and identify optimal mechanisms for specific facility characteristics and cost structures. The framework enables evaluation of coordination policies that would be infeasible to test experimentally.

More generally, our findings establish that infrastructure policies ignoring spatial interdependencies will systematically mistime interventions. Independent replacement optimization—treating each asset in isolation—proves suboptimal in networked environments where proximity creates coordination opportunities. These opportunities vary systematically with environmental stresses, implying that coordination investments should be spatially targeted rather than uniformly deployed. Organizations leaving coordination benefits unrealized forgo welfare gains that our estimates suggest are economically substantial.

\subsection{Limitations}

Several limitations qualify our findings and suggest important directions for refinement. We discuss these candidly to guide interpretation and future work.

\paragraph{Statistical Inference} Standard errors reported throughout are preliminary estimates from the numerical Hessian evaluated at the optimum. Bootstrap procedures with appropriate spatial clustering at the cage level remain incomplete, implying that all statistical inference should be interpreted cautiously. We cannot yet formally test whether coordination effects differ significantly across thermal environments, nor can we provide confidence intervals for welfare calculations. Completion of bootstrap inference represents an immediate priority for the final version.

\paragraph{Robustness Checks} Several natural robustness exercises have not yet been completed. Alternative neighborhood definitions—including distance-based specifications, hierarchical structures (cabinet-level beyond cages), and varying neighborhood radii—would verify that results are not artifacts of the specific cage-based definition. Sample period stability checks across early, middle, and late subperiods would confirm that coordination patterns reflect persistent operational behavior rather than transient phenomena. Estimation of fully interacted thermal-coordination specifications (allowing cage-specific $\gamma$ parameters) would provide direct estimates of heterogeneous effects rather than relying on moment decomposition. These analyses are planned but not yet executed.

\paragraph{Modeling Choices} Our infinite horizon formulation provides computational tractability but imposes structure inappropriate for Titan's finite operational lifetime. The system operated only seven years before decommissioning, suggesting that agents may have anticipated the terminal period and adjusted replacement timing accordingly. The infinite horizon assumption simplifies dynamic programming by eliminating terminal value functions and ensuring stationarity, but it misses end-of-life effects that likely influenced decisions in later periods. We address this limitation partially by excluding the terminal period (t = 21) from estimation, but boundary effects may affect earlier periods as well.

The asymmetric correlation moment (M2)—measuring the difference between forward and backward temporal correlation in replacement decisions—provides a diagnostic for this misspecification. In infinite horizon equilibrium, time-symmetry implies Corr(d$_{it}$, d$_{jt+1}$) = Corr(d$_{it}$, d$_{jt-1}$), yielding M2 $\approx$ 0. In finite horizon settings, temporal asymmetry emerges near the terminal period as agents accelerate or delay replacements in anticipation of shutdown. We compute M2 for validation and observe values near zero in middle periods (confirming the equilibrium approximation), but we do not include M2 in the estimation objective because doing so would introduce bias from misspecified terminal conditions. Future work incorporating finite horizon methods with explicit terminal value functions would properly model system decommissioning and exploit M2 for additional identifying variation.

Additional modeling choices warrant acknowledgment. We treat replacement as a binary decision without explicitly modeling repair as a distinct option, though our failure indicator combines events that might receive different maintenance responses. The discount factor ($\beta = 0.9$) is imposed rather than estimated, chosen to reflect quarterly discounting at conventional annual rates but potentially misspecified. Our neighborhood definition based on cage structure, while motivated by operational realities and thermal environment boundaries, remains one of several plausible specifications.

\paragraph{Data Constraints} The absence of computational workload data represents a meaningful limitation with particular relevance for extensions to AI infrastructure. We observe only time in system, not utilization intensity—each GPU's computational load (idle time versus full utilization) remains unknown. Wear patterns likely vary substantially by workload, not merely by age, with implications for both failure risks and optimal replacement timing. This limitation is particularly salient for AI infrastructure applications where workloads differ dramatically: training workloads impose intense, continuous computational demands while inference serving involves bursty, variable utilization. Our framework could incorporate utilization as an additional state variable given appropriate data, enabling dynamic workload allocation optimization. Such optimization would consider both thermal environment AND neighbor utilization patterns in placement decisions. Our findings on thermal heterogeneity suggest that workload placement matters through the coordination channel, not only through individual failure rates. The core coordination mechanisms identified here would remain relevant even with richer workload data, but predictions would strengthen and policy recommendations could become more precise.

We also lack direct cost data, limiting welfare calculations to utility terms rather than dollar magnitudes. Coordination costs—the operational expenses of synchronizing maintenance, the transaction costs of information sharing—are not directly observed but rather implicitly captured in the reduced-form spatial parameters. Detailed maintenance logs that would distinguish repair types, maintenance intensity, and component-level interventions are unavailable. Finally, our sample restriction to "normal operations" periods (excluding warranty-driven mass replacements and coordinated refresh cycles) is necessary for identification but limits external validity to routine operational decisions.

\paragraph{External Validity} Several features of our empirical setting constrain generalization. Titan represents a single facility with unique institutional characteristics. Government operation may differ from private sector environments in terms of budget constraints, incentive structures, and replacement decision authority. The homogeneous equipment deployment—all GPUs identical at system installation—proves valuable for identification but limits relevance for heterogeneous fleets with mixed vintages and multiple vendors. The predetermined spatial configuration, while ideal for isolating coordination effects, is not representative of settings where spatial relationships emerge endogenously through strategic sorting. Whether our quantitative findings on coordination magnitudes generalize to other equipment types, operational contexts, or organizational forms remains an open empirical question.

\subsection{Future Research Directions}

The framework developed here opens several promising research directions spanning methodological refinement, empirical extension, and policy application.

\paragraph{Immediate Empirical Priorities} Completing the robustness exercises discussed above represents the most direct next step. Estimation on alternative neighborhood definitions (distance-based, hierarchical cabinet-level structures, varying radii) would verify that coordination effects are not artifacts of the cage-based specification. Temporal stability analysis across early, middle, and late subperiods would confirm that spatial patterns reflect equilibrium behavior rather than transitional dynamics. Specification of cage-specific coordination parameters would provide direct estimates of thermal heterogeneity rather than relying on moment decomposition to reveal underlying variation.

\paragraph{Methodological Extensions} Several methodological advances would strengthen the framework and expand its applicability. First, developing a finite horizon formulation with explicit terminal value functions would properly model system decommissioning and enable use of asymmetric correlation moments for identification. This extension requires solving for optimal policies as functions of time-to-decommissioning, substantially increasing computational burden but eliminating the misspecification inherent in infinite horizon approximations for finite operational lifetimes.

Second, incorporating heterogeneous workload modeling would address a key data limitation. Given utilization intensity data, the framework could distinguish thermal effects from usage effects, model differential wear patterns by computational load type, and inform dynamic workload allocation decisions. This extension proves particularly valuable for AI infrastructure applications where workload heterogeneity is pronounced.

Third, developing tests for coordination optimality would distinguish observed coordination from efficient coordination. Our estimates reveal that operators do coordinate, but whether they coordinate optimally—whether observed timing and intensity maximize welfare—remains an open question. Answering this question requires characterizing the efficient coordination frontier and testing whether estimated behavior approaches it, potentially through mechanism design frameworks or comparative analysis with theoretically optimal policies.

Finally, modeling endogenous network formation would capture how coordination relationships evolve. In our setting, spatial neighbors are predetermined by facility layout. In many contexts, however, coordination relationships form endogenously as agents learn about coordination benefits, develop trust and communication channels, and strategically position themselves relative to potential coordination partners. Dynamic network formation models would capture these feedback effects.

\paragraph{Empirical Applications} The framework applies naturally to traditional infrastructure contexts beyond computing equipment. Retail chain maintenance scheduling (HVAC systems, refrigeration units, point-of-sale equipment across hundreds of stores), manufacturing facility equipment refresh cycles (production machinery, tooling, quality control systems), transportation fleet management (aircraft across hubs, vessels across ports, vehicles across depots), and distributed energy infrastructure (transformers across power grids, wind turbines across farms, solar arrays across installations) all involve spatially distributed assets where proximity creates coordination opportunities. Testing for coordination in these settings would reveal whether our findings generalize across equipment types and organizational contexts, while providing industry-specific quantitative estimates to guide operational decisions.

Emerging AI and computing infrastructure applications represent particularly promising extensions given the GPU context and rapid AI infrastructure buildout. Our framework could inform several coordination challenges in this domain. For LLM training workload allocation across distributed infrastructure, the key insight is that coordination need not be purely reactive to failures—agents can coordinate proactively based on neighbor signals that predict elevated risk. If neighbors show elevated error rates or thermal stress, operators can migrate training workloads preemptively before failures materialize, balancing migration costs against expected failure losses. This forward-looking reallocation based on degradation signals mirrors the sequential coordination we document: operators replace equipment not only when it fails but also when neighbor failures signal elevated own-unit risk.

For dynamic risk mitigation in inference serving, spatial patterns can predict failure clusters. Our finding that thermal correlations generate 10-fold coordination gradients suggests that routing traffic away from high-risk zones before failures occur may substantially improve reliability. The framework quantifies the value of spatial monitoring for predictive optimization, identifying which coordination mechanisms matter (proactive versus reactive) and where coordination generates the greatest returns (high-risk zones). Strategic spare capacity placement would prioritize regions exhibiting strong coordination benefits, informed by the spatial heterogeneity we document.

For GPU cluster management in hyperscale operations more broadly, our framework provides a template for quantifying coordination benefits and designing optimal mechanisms. The key connection is that forward-looking agents coordinate based on signals, not only events: we show neighbor failures reveal information about own future risk, and the same logic implies that neighbor degradation signals inform workload allocation and maintenance scheduling decisions. The structural parameters enable counterfactual evaluation of alternative coordination protocols—targeted early warning systems, batch maintenance scheduling, predictive replacement triggers—that would be costly or infeasible to test experimentally.

Testing for coordination in settings with endogenous sorting represents another valuable extension. How do coordination patterns differ when spatial configuration is chosen strategically versus predetermined by institutional constraints? Comparing government and private sector operations would reveal whether organizational form systematically affects coordination intensity. Analyzing heterogeneous equipment portfolios with mixed vintages and multiple vendors would test whether coordination effects vary with fleet composition.

\subsection{Concluding Remarks}

Spatial coordination in equipment replacement decisions is economically significant, systematically heterogeneous, and fundamentally shapes operational costs in networked infrastructure. Our findings establish that coordination accounts for 5.3\% of variation unexplained by independent-decision models, concentrated in high-risk environments where it matters most. Sequential coordination ($\gamma_{\text{lag}} = -0.793$) dominates reactive batching ($\gamma_{\text{fail}} = -0.265$) by 3:1, revealing that operators engage in deliberate strategic behavior. Thermal environment effects dominate chronological aging, with location-based factors generating utility penalties 34 to 47 times larger than aging effects.

These findings carry broader implications for infrastructure economics. Spatial interdependencies represent first-order considerations, not second-order refinements. Independent optimization proves systematically suboptimal in networked systems where proximity creates coordination opportunities that forward-looking agents exploit. Coordination benefits vary dramatically across risk environments—10-fold in our setting—implying that uniform policies forgo substantial welfare gains relative to spatially targeted interventions. Facility design affects operations through multiple channels: direct effects on failure rates and indirect effects through coordination constraints and opportunities.

Methodologically, the hybrid NFXP-MSM framework demonstrates that structural estimation in spatial settings need not sacrifice either tractability or interpretation. The template developed here applies broadly wherever spatial coordination shapes forward-looking decisions: retail networks, manufacturing systems, transportation fleets, distributed energy infrastructure, and emerging AI computing platforms. As computational resources scale and infrastructure becomes increasingly networked, understanding when and where coordination matters enables better mechanism design and more efficient resource allocation.

The fundamental insight is that proximity creates coordination opportunities, these opportunities vary systematically with environmental risks, and policies ignoring spatial structure leave welfare gains unrealized. Forward-looking agents in networked environments coordinate strategically, responding not only to observed failures but also to neighbor signals that predict future risks. Infrastructure management requires dynamic spatial modeling to capture these coordination effects, quantify their heterogeneity, and design policies that exploit rather than ignore the interdependencies inherent in distributed systems.

\FloatBarrier

\printbibliography

\newpage

\appendix

\section{Computational Algorithms and Performance}
\label{app:algorithms}

This appendix provides detailed algorithmic specifications for the hybrid NFXP-MSM estimation procedure, pseudocode for implementation, and computational performance metrics from the production run.

\subsection{Algorithm Overview}

The estimation procedure combines two nested optimization routines:
\begin{enumerate}
\item \textbf{Inner loop (NFXP):} For each candidate parameter vector $(\theta, \gamma)$, solve the dynamic programming problem via value function iteration to obtain choice probabilities $P(d_{it}|s_{it}; \theta, \gamma)$.
\item \textbf{Outer loop (Hybrid objective):} Evaluate the combined NFXP-MSM objective function and search over parameter space using derivative-free optimization.
\end{enumerate}

The hybrid objective function combines maximum likelihood (from NFXP) with simulated moment matching (MSM):
\begin{equation}
Q(\theta, \gamma) = -\mathcal{L}(\theta, \gamma | \text{data}) + \lambda \cdot D_{\text{MSM}}(\theta, \gamma)
\end{equation}
where $\mathcal{L}$ is the log-likelihood, $D_{\text{MSM}}$ measures distance between data and simulated moments, and $\lambda = 5.0$ is the penalty weight.

\newpage
\subsection{State Space and Transitions}

\paragraph{State Variables}
The state vector $s_{it} = (\text{age}_{it}, \text{cage}_i, \text{fail}_{it}, n^{\text{lag}}_{it})$ consists of:
\begin{itemize}
\item $\text{age}_{it} \in \{0, 1, 2, 3, 4, 5\}$: Equipment age in years (5+ pooled)
\item $\text{cage}_i \in \{0, 1, 2\}$: Thermal environment (time-invariant)
\item $\text{fail}_{it} \in \{0, 1\}$: Current period failure indicator
\item $n^{\text{lag}}_{it} \in \{0, 1\}$: At least one neighbor replaced in $t-1$
\end{itemize}

The transient spatial variable $f^{\text{cage}}_{it} \in \mathbb{Z}_+$ (neighbor failure count) enters flow utility but not the persistent state space.

Total state space dimension: $6 \times 3 \times 2 \times 2 = 72$ states.

\paragraph{State Transitions}
Age transitions deterministically upon continuation:
\begin{equation}
\text{age}_{i,t+1} = \begin{cases}
0 & \text{if } d_{it} = 1 \text{ (replace)} \\
\min(\text{age}_{it} + \Delta t, 5) & \text{if } d_{it} = 0 \text{ (keep)}
\end{cases}
\end{equation}
where $\Delta t = 0.25$ years (quarterly time step).

Thermal environment is time-invariant: $\text{cage}_{i,t+1} = \text{cage}_i$.

Failure transitions follow empirical probabilities estimated from observed state transitions with Laplace smoothing ($\alpha = 0.01$):
\begin{equation}
P(\text{fail}_{i,t+1} = k' | \text{age}_{it}, \text{cage}_i, \text{fail}_{it} = k, d_{it}) = \frac{N_{k \to k'} + \alpha}{N_k + 2\alpha}
\end{equation}

Replacement deterministically resets failure status: $P(\text{fail}_{i,t+1} = 0 | d_{it} = 1) = 1$.

Neighbor replacement indicator evolves based on neighbor actions:
\begin{equation}
n^{\text{lag}}_{i,t+1} = \mathbb{I}\left[\sum_{j \in \mathcal{N}(i)} d_{jt} \geq 1\right]
\end{equation}

Expected neighbor replacement probability $p^{\text{nbr}}$ is computed as the unconditional replacement rate in the data.

\subsection{Value Function Iteration (NFXP Inner Loop)}

\begin{algorithm}[H]
\caption{Value Function Iteration}
\begin{algorithmic}[1]
\STATE \textbf{Input:} Parameters $(\theta, \gamma)$, discount factor $\beta = 0.9$, tolerance $\epsilon = 10^{-4}$
\STATE \textbf{Initialize:} $V^0(s) = 0$ for all states $s$
\STATE Set iteration counter $k = 0$
\REPEAT
    \STATE $k \leftarrow k + 1$
    \FOR{each state $s = (\text{age}, \text{cage}, \text{fail}, n^{\text{lag}})$}
        \STATE Compute flow utilities:
        \STATE \quad $u_{\text{keep}}(s) = \theta_{\text{age}} \cdot \text{age} + \theta_{\text{cage}_1} \cdot \mathbb{I}[\text{cage}=1] + \theta_{\text{cage}_2} \cdot \mathbb{I}[\text{cage}=2]$
        \STATE \quad\quad $+ \theta_{\text{fail}} \cdot \text{fail} + \gamma_{\text{lag}} \cdot n^{\text{lag}} + \gamma_{\text{fail}} \cdot \mathbb{E}[f^{\text{cage}}]$
        \STATE \quad $u_{\text{replace}}(s) = \theta_{\text{replace}}$
        \STATE Compute expected continuation values:
        \STATE \quad $EV_{\text{keep}}(s) = \sum_{s'} P(s'|s, d=0) \left[ p^{\text{nbr}} V^{k-1}(s', n^{\text{lag}}=1) + (1-p^{\text{nbr}}) V^{k-1}(s', n^{\text{lag}}=0) \right]$
        \STATE \quad $EV_{\text{replace}}(s) = \sum_{s'} P(s'|s, d=1) \left[ p^{\text{nbr}} V^{k-1}(s', n^{\text{lag}}=1) + (1-p^{\text{nbr}}) V^{k-1}(s', n^{\text{lag}}=0) \right]$
        \STATE Update value function (Bellman operator):
        \STATE \quad $V^k(s) = \log\left[ \exp(u_{\text{keep}}(s) + \beta \cdot EV_{\text{keep}}(s)) + \exp(u_{\text{replace}}(s) + \beta \cdot EV_{\text{replace}}(s)) \right]$
    \ENDFOR
    \STATE Compute convergence metric: $\Delta = \max_s |V^k(s) - V^{k-1}(s)|$
\UNTIL{$\Delta < \epsilon$ or $k > 2000$}
\STATE \textbf{Output:} Converged value function $V^*(s) = V^k(s)$
\end{algorithmic}
\end{algorithm}

\paragraph{Choice Probabilities}
Given the converged value function $V^*(s)$, choice probabilities follow the logit form:
\begin{equation}
P(d_{it} = 1 | s_{it}; \theta, \gamma) = \frac{\exp\left( u_{\text{replace}}(s_{it}) + \beta \cdot EV_{\text{replace}}(s_{it}) \right)}{\exp\left( u_{\text{keep}}(s_{it}) + \beta \cdot EV_{\text{keep}}(s_{it}) \right) + \exp\left( u_{\text{replace}}(s_{it}) + \beta \cdot EV_{\text{replace}}(s_{it}) \right)}
\end{equation}

\newpage
\subsection{Method of Simulated Moments (MSM Component)}

\paragraph{Moment Definitions}
We target three spatial moments that isolate coordination effects:

\textbf{M1 (Sequential hazard):} Replacement persistence after own prior replacement
\begin{equation}
m_1 = \mathbb{E}[d_{it} | d_{i,t-1} = 1]
\end{equation}

\textbf{M3 (Lagged failure spillover):} Replacement response to neighbor failures in $t-1$, among non-failed units
\begin{equation}
m_3 = \mathbb{E}[d_{it} | \text{fail}_{it} = 0, f^{\text{nbr}}_{i,t-1} \geq 1] - \mathbb{E}[d_{it} | \text{fail}_{it} = 0, f^{\text{nbr}}_{i,t-1} = 0]
\end{equation}

\textbf{M4 (Current failure batching):} Replacement response to contemporaneous neighbor failures, among non-failed units
\begin{equation}
m_4 = \mathbb{E}[d_{it} | \text{fail}_{it} = 0, f^{\text{cage}}_{it} \geq 1] - \mathbb{E}[d_{it} | \text{fail}_{it} = 0, f^{\text{cage}}_{it} = 0]
\end{equation}

\paragraph{Moment Distance Function}
The MSM objective uses unweighted squared deviations:
\begin{equation}
D_{\text{MSM}}(\theta, \gamma) = \sum_{j \in \{1,3,4\}} \left( m_j^{\text{data}} - m_j^{\text{sim}}(\theta, \gamma) \right)^2
\end{equation}

\paragraph{Forward Simulation}
Simulated moments are computed by forward-simulating $S = 50$ panels of length $T = 13$ quarters:
\begin{enumerate}
\item Initialize each panel at observed period-8 states from the data
\item For each period $t$:
    \begin{itemize}
    \item Draw choice probabilities $P(d_{it}=1|s_{it}; \theta, \gamma)$ from NFXP solution
    \item Sample binary replacement decisions: $d_{it} \sim \text{Bernoulli}(P(d_{it}=1|s_{it}))$
    \item Update states according to transition probabilities
    \item Record neighbor replacement indicators and failure counts
    \end{itemize}
\item Compute moments $m_j^{\text{sim}}$ across all simulated observations
\item Average across $S$ simulation draws
\end{enumerate}

\newpage
\subsection{Outer Loop Optimization}

\begin{algorithm}[H]
\caption{Hybrid NFXP-MSM Estimation}
\begin{algorithmic}[1]
\STATE \textbf{Input:} Data $\{d_{it}, s_{it}\}$, penalty weight $\lambda = 5.0$, simulation draws $S = 50$
\STATE \textbf{Initialize:} Parameter vector $(\theta^0, \gamma^0)$ from reduced-form estimates
\STATE Set optimization method: Nelder-Mead simplex algorithm
\REPEAT
    \STATE Receive candidate parameters $(\theta, \gamma)$ from optimizer
    \STATE \textbf{NFXP Step:}
    \STATE \quad Solve value function iteration (Algorithm 1) → obtain $V^*(s; \theta, \gamma)$
    \STATE \quad Compute choice probabilities $P(d_{it}=1|s_{it}; \theta, \gamma)$
    \STATE \quad Evaluate log-likelihood: $\mathcal{L}(\theta, \gamma) = \sum_{i,t} \left[ d_{it} \log P(d_{it}=1|s_{it}) + (1-d_{it}) \log P(d_{it}=0|s_{it}) \right]$
    \STATE \textbf{MSM Step:}
    \STATE \quad Forward-simulate $S$ panels using $P(d_{it}=1|s_{it}; \theta, \gamma)$
    \STATE \quad Compute simulated moments $\{m_1^{\text{sim}}, m_3^{\text{sim}}, m_4^{\text{sim}}\}$
    \STATE \quad Evaluate moment distance: $D_{\text{MSM}}(\theta, \gamma) = \sum_j (m_j^{\text{data}} - m_j^{\text{sim}})^2$
    \STATE \textbf{Objective:}
    \STATE \quad $Q(\theta, \gamma) = -\mathcal{L}(\theta, \gamma) + \lambda \cdot D_{\text{MSM}}(\theta, \gamma)$
    \STATE Return $Q(\theta, \gamma)$ to optimizer
\UNTIL{Convergence criterion met or maximum iterations reached}
\STATE \textbf{Output:} Estimated parameters $(\hat{\theta}, \hat{\gamma})$
\end{algorithmic}
\end{algorithm}

\paragraph{Optimization Details}
\begin{itemize}
\item \textbf{Method:} Nelder-Mead simplex algorithm (derivative-free)
\item \textbf{Rationale:} Objective function non-smooth due to simulation noise; gradient-based methods unstable
\item \textbf{Convergence:} Adaptive step sizes with tolerance on parameter changes and objective improvement
\item \textbf{Warm starts:} Value function initialized from previous iteration to reduce computational burden
\end{itemize}

\newpage
\subsection{Computational Performance}

Table \ref{tab:comp_performance} reports computational metrics from the production estimation run on the full sample (147,078 observations, 12,915 locations, 13 time periods).

\begin{table}[H]
\centering
\caption{Computational Performance Metrics}
\label{tab:comp_performance}
\begin{tabular}{lc}
\toprule
\textbf{Metric} & \textbf{Value} \\
\midrule
\multicolumn{2}{l}{\textit{Problem Dimensions}} \\
State space size & 72 states \\
Observations & 147,078 \\
Parameters estimated & 7 \\[6pt]
\multicolumn{2}{l}{\textit{Algorithm Settings}} \\
Discount factor ($\beta$) & 0.90 \\
VF iteration tolerance ($\epsilon$) & $10^{-4}$ \\
Max VF iterations & 2,000 \\
MSM simulations ($S$) & 50 \\
MSM penalty weight ($\lambda$) & 5.0 \\[6pt]
\multicolumn{2}{l}{\textit{Estimation Results}} \\
Final objective value & 6,123.75 \\
Log-likelihood & $-6,123.75$ \\
Outer loop iterations & 295 \\
Function evaluations & 295 \\
Converged & No \\[6pt]
\multicolumn{2}{l}{\textit{Computation Time}} \\
Total elapsed time & 45.9 hours \\
Average time per iteration & 9.3 minutes \\
\bottomrule
\end{tabular}
\begin{minipage}{\textwidth}
\vspace{6pt}
\footnotesize
\textit{Notes:} Production run executed sequentially. Formal 
convergence criterion not met within iteration limit. However, informal 
monitoring during estimation indicated parameter stability in later iterations, suggesting the algorithm approached a local optimum. Total time includes NFXP solution and MSM component for each iteration. Average time calculated as 
total elapsed time divided by iterations.
\end{minipage}
\end{table}

\paragraph{Computational Bottlenecks}
The primary computational burden arises from:
\begin{enumerate}
\item \textbf{Value function iteration:} Solved from scratch for each candidate parameter vector during outer loop optimization.
\item \textbf{Forward simulation:} Multiple simulation draws across all locations and time periods for MSM moment calculation.
\item \textbf{Outer loop:} 295 function evaluations, each requiring complete NFXP solution plus MSM component.
\end{enumerate}

\begin{samepage}
\paragraph{Computational Implementation}
The current implementation executes sequentially with the following optimizations:
\begin{itemize}
\item \textbf{Value function vectorization:} State space operations vectorized rather than looped
\item \textbf{Warm starts:} Initial value function from previous iteration reduces convergence time
\item \textbf{Cached transitions:} Empirical transition probabilities pre-computed and stored
\end{itemize}
\end{samepage}
\paragraph{Potential Parallelization}
Substantial computational speedups could be achieved through parallelization:
\begin{itemize}
\item \textbf{Simulation parallelization:} The 50 MSM simulation draws are independent and could be distributed across processor cores, potentially reducing MSM computation time by a factor of 10-50×
\item \textbf{Spatial domain decomposition:} Moment calculations across locations could be parallelized
\item \textbf{Parameter grid search:} Multiple candidate parameter vectors could be evaluated simultaneously during outer loop optimization
\end{itemize}
These parallelization opportunities represent promising directions for reducing computation time in future implementations.

\newpage
\subsection{Parameter Initialization}

Initial parameter values for outer loop optimization derived from reduced-form estimates:
\begin{itemize}
\item \textbf{Structural parameters ($\theta$):} Logit regression of replacement on age, cage, failure
\item \textbf{Spatial parameters ($\gamma$):} Coefficients on $n^{\text{lag}}_{it}$ and $f^{\text{cage}}_{it}$ from augmented logit
\item \textbf{Scaling:} Initial values scaled to match approximate magnitude of structural utility parameters
\end{itemize}

This initialization strategy ensures the optimizer begins in a plausible region of parameter space, reducing computation time and improving convergence reliability.

\subsection{Robustness and Diagnostics}

\paragraph{Convergence Status}
The optimization routine completed 295 iterations without meeting formal 
convergence criteria. However, informal monitoring during estimation indicated 
that parameter values stabilized in later iterations, suggesting proximity to 
a local optimum. The estimated parameters are consistent with economic theory 
(negative coefficients for age, thermal stress, and failure effects) and 
generate spatial moments closely matching observed data patterns, as documented 
in Table \ref{tab:moment_validation}.

\paragraph{Computational Feasibility}
The hybrid NFXP-MSM approach proves computationally tractable for this high-dimensional spatial problem through parallelization of simulation draws and vectorization of state space operations. Total computation time of 45.9 hours (approximately 2 days) is feasible for research applications, though further optimization of the value function iteration routine could reduce runtime.

\FloatBarrier

\end{document}